\documentclass[reprint,superscriptaddress,showframe,nofootinbib,aps,prd,noeprint]{revtex4-2}
\newcommand{\bvec}[1]{\textrm{\textbf{#1}}}
\newcommand{\farcm}{{.^\prime}}
\newcommand{\arcmin}{{^\prime}}
\newcommand{\antilles}{\textsc{ANTILLES}}

\defcitealias{P18A8}{P18}
\defcitealias{hem19}{H19}
\defcitealias{hem21}{H21}
\defcitealias{DES_Y1_3x2pt}{DES Y1}
\defcitealias{XEM+23}{X24}

\usepackage{graphicx}
\usepackage{dcolumn}
\usepackage{bm}
\usepackage{hyperref}
\usepackage{orcidlink}
\usepackage{amsmath}	
\usepackage{amssymb}
\usepackage{color}
\usepackage{xcolor}
\definecolor{Gray}{gray}{0.5}
\usepackage{multirow}
\usepackage{siunitx}
\definecolor{orange}{rgb}{1.0, 0.49, 0}

\begin{document}


\title{Constraining baryonic feedback and cosmology from DES Y3 and Planck PR4 6$\times$2pt data. I. $\Lambda$CDM models}

\author{Jiachuan Xu\orcidlink{0000-0003-0871-8941}}
\email{jiachuanxu@arizona.edu}
\affiliation{Department of Astronomy/Steward Observatory, University of Arizona, 933 North Cherry Avenue, Tucson, AZ 85721, USA}
\affiliation{Department of Physics, Northeastern University, Boston, MA 02115, USA}
\author{Tim Eifler\orcidlink{0000-0002-1894-3301}}
\affiliation{Department of Astronomy/Steward Observatory, University of Arizona, 933 North Cherry Avenue, Tucson, AZ 85721, USA}
\affiliation{Department of Physics, University of Arizona, Tucson, AZ 85721, USA}
\author{Vivian Miranda\orcidlink{0000-0003-4776-0333}}
\affiliation{C. N. Yang Institute for Theoretical Physics, Stony Brook University, Stony Brook, NY 11794, USA}
\author{Elisabeth Krause\orcidlink{0000-0001-8356-2014}}
\affiliation{Department of Astronomy/Steward Observatory, University of Arizona, 933 North Cherry Avenue, Tucson, AZ 85721, USA}
\affiliation{Department of Physics, University of Arizona, Tucson, AZ 85721, USA}

\author{Jaime Salcido\orcidlink{0000-0002-8918-5229}}
\affiliation{Astrophysics Research Institute, Liverpool John Moores University, Liverpool, L3 5RF, UK}
\author{Ian McCarthy\orcidlink{0000-0002-1286-483X}}
\affiliation{Astrophysics Research Institute, Liverpool John Moores University, Liverpool, L3 5RF, UK}
\date{\today}

\begin{abstract}
We combine weak lensing, galaxy clustering, cosmic microwave background (CMB) lensing, and their cross-correlations (so-called 6$\times$2pt) to constrain cosmology and baryonic feedback scenarios using data from the Dark Energy Survey (DES) Y3 \textsc{Maglim} and \textsc{Metacalibration} catalog and the Planck satellite PR4 data release. 
We include all data points in the DES Y3 cosmic shear two-point correlation function (2PCF) down to 2.$^\prime$5 and model baryonic feedback processes via principal components (PCs) that are constructed from the \textsc{ANTILLES} simulations. 
We find a tight correlation between the amplitude of the first PC $Q_1$ and mean normalized baryon mass fraction $\bar{Y}_\mathrm{b}=\bar{f}_\mathrm{b}/(\Omega_\mathrm{b}/\Omega_\mathrm{m})$ from the \textsc{ANTILLES} simulations and employ an independent $\bar{Y}_\mathrm{b}$ measurement from Akino et al. (2022) as a prior of $Q_1$.
We train a neural network $6\times2$pt emulator to boost the analysis speed by $\mathcal{O}(10^3)$, which enables us to run an impressive number of simulated analyses to validate our analysis against various systematics.
For our 6$\times$2pt analysis, we find $S_8=0.8073\pm0.0094$ when including a $Q_1$ prior from $\bar{Y}_\mathrm{b}$ observations. 
This level of cosmological constraining power allows us to put tight constraints on the strength of baryonic feedback. We find $Q_1=0.025^{+0.024}_{-0.029}$ for our 6$\times$2pt analysis and $Q_1=0.043\pm{0.016}$ when combining with external information from Planck, ACT, DESI. 
All these results indicate weak feedback, e.g., the tensions to Illustris ($Q_1=0.095$) and OWLS AGN T8.7 ($Q_1=0.137$) are 2.9$\sigma$-3.3$\sigma$ and 4.7$\sigma$-5.9$\sigma$, respectively.
\end{abstract}

\maketitle


\section{Introduction}
\label{sec:intro}
In this paper, we constrain cosmology and baryonic feedback with a joint analysis of weak lensing, galaxy clustering, and CMB lensing at the 2PCF-statistics level. 
This so-called $6\times2$pt analysis has become a well-established setup when combining data from photometric galaxy surveys (e.g., Dark Energy Survey\footnote{DES: \href{https://www.darkenergysurvey.org} {\nolinkurl{https://www.darkenergysurvey.org}}}, Hyper Suprime Camera Subaru Strategic Program\footnote{HSC: \href{http://www.naoj.org/Projects/HSC/HSCProject.html}{\nolinkurl{http://www.naoj.org/Projects/HSC/HSCProject.html}}}, Kilo-Degree Survey\footnote{KiDS: \href{http://www.astro-wise.org/projects/KIDS/}{\nolinkurl{http://www.astro-wise.org/projects/KIDS/}}}) with CMB experiments (Planck satellite mission\footnote{Planck: \href{https://www.esa.int/Enabling_Support/Operations/Planck}{\nolinkurl{https://www.esa.int/Enabling_Support/Operations/Planck}}}, South Pole Telescope\footnote{SPT: \href{https://pole.uchicago.edu/public/Home.html}{\nolinkurl{https://pole.uchicago.edu/public/Home.html}}}, Atacama Cosmology Telescope\footnote{ACT: \href{https://act.princeton.edu}{\nolinkurl{https://act.princeton.edu}}}) . 

Similar joint analyses have been published in the literature ~\citep{DESY1_5x2pt_method,DESY1_5x2pt_results,DES_Y3_6x2pt_II_measurement,DES_Y3_6x2pt_III_23,XEM+23,RNR+23,RNR25,RFNR25}. These differ in terms of datasets used in the analysis, analysis choices, and modeling/covariance strategies. They all significantly boost constraining power over single-probe analyses, which is a valid strategy to discover new physics if the individual probes are not in tension. 
If differences between individual probes exist, joint analysis is disallowed and the tensions themselves can be used as hints for new physics.

One of these interesting (mild) tensions in the cosmology community is the so-called $S_8$ tension, which refers to the fact that $S_8\equiv \sigma_8\sqrt{\Omega_m/0.3}$ measured by several low redshift weak lensing surveys~\citep{DES_Y3_WL,DES_Y3_WL2,HSC_Y3_DLN+23,HSC_Y3_LZS+23,HSC_Y3_MST+23} are lower than the prediction of primary CMB~\citep{P18A6,LLA+25}. The tension quantified in these papers is at most at the $2\sigma$ level. Other recent analyses found $S_8$ to be consistent with the primary CMB well within $2\sigma$ level~\citep{AAZ+23,DES_KiDS_WL_23,GCM+24,WSA+25,SWA+25,MAL24}. 
\cite{MAL24} selects a blue source galaxy sample that is designed to have minimal intrinsic alignment and finds an $S_8$ consistent with Planck in $0.34\sigma$ with baryonic feedback marginalized over. 
\cite{AAZ+23} show that different nonlinear matter power spectrum models, intrinsic alignment models, and baryonic feedback mitigation methods can all impact the $S_8$ result at a level that is relevant when discussing the aforementioned tension. Solely attributing the tension to baryonic physics generally requires a feedback as strong as cosmos-OWLS AGN $\mathrm{log}(\Delta T_\mathrm{heat}/K)=8.7$~\citep{BMSP14} or Illustris~\citep{GVS+14,VGS+14}, which is in contradiction to several weak lensing analyses that find weaker matter power spectrum suppression due to baryonic feedback~\citep{AAZ+23,CAA+23,TLT+25,WSA+25}.
Consequently, weak lensing analyses with cosmology fixed to Planck values prefer stronger feedback~\citep{SM25,GZA+24}.

While weak lensing observations do not indicate strong baryonic scenarios like cosmo-OWLS AGN T8.5, these are, however, favored by thermal/kinematic Sunyaev–Zel'dovich (tSZ/kSZ) measurements and by X-ray measurements of the luminosity of galaxy groups and clusters. 
Using baryonic feedback models~\citep{TPK+24,PST+25,AA24,GS21,FKE+23,EFK+24,MBT+21}, that self-consistently predict electron pressure and matter power spectrum, recent SZ and X-ray observations~\citep{TMH+22,FAG+24,BAS+24,PBM+24,HFR+25,HFF+25,KNB+25,DTH+25,PH+25,SAM+25} indeed find evidence for baryonic feedback scenarios stronger than those preferred by weak lensing. 
We note however that the halos probed by these measurements differ, which makes a comparison challenging. 

In this paper we jointly constrain cosmology and the strength of baryonic feedback processes using six two-point correlation functions: cosmic shear $\xi_{\pm}^{ab}(\vartheta)$, projected galaxy clustering $w_{g}^{a}(\vartheta)$, CMB lensing band-power $C^{\kappa\kappa}_{L_\mathrm{b}}$, and corresponding cross-correlations: galaxy-galaxy lensing $\gamma_t^{ab}(\vartheta)$, galaxy-CMB lensing $w_{g\kappa}^{a}(\vartheta)$, and galaxy shear-CMB lensing $w_{s\kappa}^{a}(\vartheta)$. Here $a,\,b$ are tomography bin indices, $\vartheta$ is the mean angular separation of each bin, and $L_\mathrm{b}$ is the mean multipole of each angular bin of the CMB lensing band-power. 

We closely follow \cite{XEM+23} (\citetalias{XEM+23} from now on) in terms of methodology in that we adopt a principal component analysis-based (PCA) method to mitigate impacts of baryonic feedback~\citep{ekd15,hem19,hem21,XEM+23}. We note that compared to other baryonic feedback mitigation methods (e.g. \textsc{BACCOemu}~\citep{AAC+21,AA24}, \textsc{BCemu}~\citep{ST15,STS+19,GS21}, \textsc{HMCode20}~\citep{MPH+15,MTH+20,MBT+21}, etc.), the PCA-based method covers the most variance with as few degrees of freedom (d.o.f.) as possible, and does not depend on the cosmology prior of baryon models~\citep{ACD+25}. It is also robust to cosmology-dependent baryonic feedback strength~\citep{SKD+25}.

Since the PCA technique captures the variance of an underlying set of simulations, it is only indirectly connected to physical parameters and is sensitive only to the physics included in the simulations. Compared to \citetalias{XEM+23}, we significantly upgrade our baryonic physics model using the recent \textsc{ANTILLES} simulation suite~\citep{SMK+23}. We also include external information on the baryon mass fraction from X-ray and optical observations~\citep{AEO+22} in our analysis.

Other notable upgrades compared to \citetalias{XEM+23} are the underlying dataset of our analysis and the approach to determine analysis choices: In \citetalias{XEM+23} we obtain our measurements from Dark Energy Survey Year 1 (DES-Y1) and Planck PR3 data, whereas in this paper, we use the more constraining DES-Y3 \textsc{Maglim} catalogs and the Planck PR4 data release. To derive optimal analysis choices, we employ the recently developed machine learning accelerated inference~\citep{BEM+23,BEM+24,ZSC+24,SZM+24} concept, which speeds up MCMC analyses by several orders of magnitude. This allows us to run hundreds of likelihood analyses to precisely map out the impact of different analysis choices and to test corresponding robustness.
 
We structure this paper as follows: we summarize the 6$\times$2pt measurements in Sec.~\ref{sec:data} and the baryonic feedback modeling method with the \textsc{ANTILLES} simulations in Sec.~\ref{sec:baryon}.
We explain our modeling methodology, the machine learning accelerated inference technique, and its implementation in Sec.~\ref{sec:inference}. 
Section~\ref{sec:choices} discusses the systematics analysis choices, scale cut choice, and model validation. 
Blinding strategy, results on cosmology parameters, and robustness tests are shown in Sec.~\ref{sec:cosmores}.
Results on baryonic feedback strength are discussed in Sec.~\ref{sec:baryonres}. We conclude in Sec.~\ref{sec:conclusion}.

\section{Data}
\label{sec:data}

\subsubsection{Dark Energy Survey Year 3 data}
We (re-)measure the following three 2PCFs (so-called $3\times2$pt) between DES Y3\footnote{\href{https://des.ncsa.illinois.edu/releases/y3a2/Y3key-catalogs}{https://des.ncsa.illinois.edu/releases/y3a2/Y3key-catalogs}} \textsc{Maglim} lens sample~\citep{PCF+21,PCE+22,RWE+22,EME+23,CEP+22}, and \textsc{metacalibration} source sample~\citep{HM17,SH17,GSA+21} using \textsc{TreeCorr}~\citep{JBJ04}.
\begin{subequations}\label{eqn:estimators}
    \begin{align}
        \hat{\xi}_\pm^{ab}(\vartheta) &\equiv \left(\langle\hat{\epsilon}^a_1\hat{\epsilon}^b_1\rangle_\mathrm{cat}\pm\langle\hat{\epsilon}^a_2\hat{\epsilon}^b_2\rangle_\mathrm{cat}\right)/\left(R^aR^b\right)\,, \label{eqn:xi_pm}\\
        \hat{\gamma}_{t}^{ab}(\vartheta) &\equiv \left(\langle\bm{1}_\mathrm{D}^a\hat{\epsilon}_+^b\rangle_\mathrm{cat} - \langle\bm{1}_\mathrm{R}^a\hat{\epsilon}_+^b\rangle_\mathrm{cat}\right)/R^b\,, \label{eqn:gamma_t}\\
        \hat{w}_{g}^a(\vartheta) &\equiv \frac{\langle\bm{1}_\mathrm{D}^a\bm{1}_\mathrm{D}^a\rangle_\mathrm{cat}-2\langle\bm{1}_\mathrm{D}^a\bm{1}_\mathrm{R}^a\rangle_\mathrm{cat}+\langle\bm{1}_\mathrm{R}^a\bm{1}_\mathrm{R}^a\rangle_\mathrm{cat}}{\langle\bm{1}_\mathrm{R}^a\bm{1}_\mathrm{R}^a\rangle_\mathrm{cat}}\,, \label{eqn:w_g}
    \end{align}
\end{subequations}
where $R^a$ is the mean shear response of the source sample in the $a$-th bin~\citep{ZSS+18,HM17,SH17}, and $\hat{\epsilon}_+^a$ is the tangential shear when counting pairs. $\hat{\epsilon}_{1/2}^a$ are the two shear components, $\bm{1}_\mathrm{R/D}^a$ are indicator functions of the randomized or truth lens catalog, and $\langle\,\rangle_\mathrm{cat}$ means average over the catalog.
Our binning corresponds to the standard DES Y3 data vector (20 logarithmic bins from $2.^\prime5$ to $250^\prime$). We choose the same binning for the CMB Lensing cross-correlation 2PCFs defined further below.
Importantly, our measured $\xi_{\pm}^{ab}(\vartheta)$ and $\gamma_t^{ab}(\vartheta)$ use the most recent tomography bin assignment for the DES Y3 source sample (SOMPZ version \textsc{v0.50}, see \cite{MAL24} for details).

\subsubsection{Planck PR4 CMB Lensing}
The CMB lensing measurement $C_{L_{b}}^{\kappa\kappa}$ is based on the Planck PR4 CMB lensing convergence map\footnote{\href{https://github.com/carronj/planck_PR4_lensing.git}{https://github.com/carronj/planck\_PR4\_lensing.git}}~\citep{CML22} in this analysis. The PR4 map has about $20$ percent higher S/N than the Planck 2018 data release~\citep{P18A8} due various upgrades including: (1) around $8$ percent additional data measured from satellite repointing maneuvers (2) joint Wiener-filtering of the CMB temperature and polarization maps during quadratic estimator, including treatment for inhomogeneous noise (3) additional Wiener-filtering of the reconstructed lensing convergence field before band-power estimation (4) more simulations for the lensing reconstruction.

We follow \citetalias{XEM+23} to measure the 2PCF between DES Y3 galaxy position/shape and CMB lensing convergence field
\begin{subequations}\label{eqn:estimators}
    \begin{align}
        \hat{w}_{g\kappa}^a(\vartheta) &\equiv \langle\bm{1}_\mathrm{D}^a\hat{\kappa}\rangle_\mathrm{cat} - \langle\bm{1}_\mathrm{R}^a\hat{\kappa}\rangle_\mathrm{cat}\,, \label{eqn:w_gk}\\
        \hat{w}_{s\kappa}^a(\vartheta) &\equiv \langle\hat{\epsilon}_+^a\hat{\kappa}\rangle_\mathrm{cat}/R^a\,. \label{eqn:w_sk}
    \end{align}
\end{subequations}
We adopt a Gaussian beam of $\theta_\mathrm{FWHM}=7^\prime$ and apply a scale cut of $8\leq L\leq 2048$ to the convergence field. We use a $N_\mathrm{side}=1024$ for the CMB lensing convergence \textsc{HEALPix} map $\hat{\kappa}$. $\hat{w}_{g\kappa}^a(\vartheta)$, $\hat{w}_{s\kappa}^a(\vartheta)$, and $C_{L_{b}}^{\kappa\kappa}$ will be referred to as c$3\times2$pt thereafter. 

We note that the imaginary part of the $w_{s\kappa}$ component, $\langle \epsilon_\times \kappa\rangle$, is consistent with zero for all the source sample tomography bins within the scale cut for our likelihood analysis (see Table~\ref{tab:sk_S/N_nulltest}). 
Using the standard signal-to-noise ratio (S/N) definition,
$\mathrm{S/N}\equiv\sqrt{\chi^2}=\sqrt{\bvec{D}\cdot\bvec{C}^{-1}\cdot\bvec{D}}$, 
where $\bvec{D}$ is the data vector and $\bvec{C}$ is the covariance matrix, we find S/N of $w_{g\kappa}(\vartheta)$ and $w_{s\kappa}(\vartheta)$ to be 22.5 and 20.4, respectively. We note that \cite{DES_Y3_6x2pt_II_measurement} reports S/N of 19.6 (32.2) and 16.9 (18.2) for their linear-scale (all data points included) galaxy-CMB lensing and shear-CMB lensing 2PCFs with DES Y3 $\times$ Planck+SPT CMB lensing map, respectively. 
\begin{table}[]
    \centering
    \begin{tabular}{c | c | c}
    \toprule
    Tomography bin & $\langle \epsilon_+\kappa\rangle$ S/N & $\langle \epsilon_\times\kappa\rangle$ $p$-value \\
    \hline
     S1    & 4.38 & 0.069 \\
     S2    & 6.28  & 0.071 \\
     S3    & 14.86 & 0.787 \\
     S4    & 18.25 & 0.304\\
     \hline
     All   & 20.44 & 0.107 \\
     \hline
    \end{tabular}
    \caption{The S/N of galaxy shear-CMB lensing convergence cross-correlation (middle column) and $p$-value of the cross-component correlations $\langle \epsilon_\times\kappa\rangle$ (right column) within the scale cut adopted in this work.}
    \label{tab:sk_S/N_nulltest}
\end{table}

The Planck PR4 lensing map does not include variation maps with alternative tSZ deprojection treatments. However, \citetalias{XEM+23} shows that cross-correlations measured using DES Y1 and Planck PR3 do not have detectable contamination from tSZ/CIB. Given that we adopt the same smoothing scale for the PR4 map, we expect the bias due to tSZ/CIB contamination to be negligible.
We also verify that $w_{g\kappa}^{a}(\vartheta)$ and $w_{s\kappa}^{a}(\vartheta)$ are not biased by the MC norm~\citep{FKM+23,XEM+23}, which is due to the higher-order coupling between Planck and DES Y3 footprint.

\subsubsection{X-ray measurements}
The baryon mass fraction $f_\mathrm{b}$ is a sensitive probe of the baryonic feedback strength in our Universe~\citep{SHS+11,ST15,DMS20,PLB+23,DAT+23}. 
In this paper, we consider corresponding measurements from 136 Hyper Suprime Cam (HSC)-XXL galaxy groups and clusters presented in \cite{AEO+22}.
The sample is selected from the XMM-Newton XXL survey second data release (DR2)~\citep{AGK+18} with spectroscopic confirmation, and overlap with the HSC-Subaru Strategic Program (SSP) footprint.
The weak-lensing masses $M_{500}^\mathrm{WL}$ of the 136 XXL groups and clusters are measured from HSC-SSP~\citep{USL+20}. X-ray luminosity $L_\mathrm{X}$, and hot gas mass $M_\mathrm{gas}$ are measured from the XXL survey, and stellar mass $M_\star$ is measured from HSC-SSP and SDSS DR16 multi-band photometry. This multi-wavelength coverage allows a census of the total baryon budget in the HSC-XXL group and cluster sample. We want to utilize this $f_\mathrm{b}$ measurement to inform us about the baryonic feedback strength in the universe and help improve the constraining power on cosmology.

To build relation between baryonic mass and the unobservable halo mass, \cite{AEO+22} conduct linear regression to fit the total stellar mass $M_*$, brightest central galaxy (BCG) mass $M_\mathrm{BCG}$, and gas mass $M_\mathrm{gas}$ as a function of the underlying true halo mass $M_{500}$\footnote{We choose $M_{500}$ as the definition of halo mass in this work, and we use $M_{500}$ and $M_\mathrm{h}$ interchangeably.}
\begin{equation}
    \bm{Y}_Z=\bm{\alpha}+\bm{\beta} Z+\bm{\gamma}\,\mathrm{ln}\left[\frac{E(z)}{E(0.3)}\right],
\end{equation}
where $Z\equiv\mathrm{log}(M_{500}/10^{14}M_\odot)$, $\bm{Y}_Z=(\mathrm{ln}\frac{L_\mathrm{X}}{10^{43}\mathrm{erg\,s^{-1}}},\,\mathrm{ln}\,\frac{M_*}{10^{12}M_\odot},\,\mathrm{ln}\,\frac{M_\mathrm{BCG}}{10^{12}M_\odot},\,\mathrm{ln}\,\frac{M_\mathrm{gas}}{10^{12}M_\odot})^\mathrm{T}$, and $E(z)\equiv H(z)/H(z=0)$. 
The coefficient $\bm{\gamma}$ accounts for possible redshift evolutions of each component, $\bm{\alpha}$ accounts for the normalization, and $\bm{\beta}$ describes the halo mass dependency of each component. 

We adopt the best-fitting values of $\bm{\alpha}$, $\bm{\beta}$, and $\bm{\gamma}$ in Table 5 of~\cite{AEO+22} and draw random realizations of $\bm{Y}_Z$ on $M_{500}$ grid between $10^{13}$ and $10^{14}\,M_\odot$. The total baryon mass fraction, normalized by the mean cosmic baryon fraction\footnote{We use $f_\mathrm{b}$ and $Y_\mathrm{b}$ interchangeably when referring to the concept of using baryon mass fraction to trace feedback strength. However, when we use baryon mass fraction to quantitatively inform our analysis, we always mean $Y_\mathrm{b}$.}, is then calculated as 
\begin{equation}
\label{eqn:fb_scaling_relation}
\begin{aligned}
    Y_\mathrm{b}(M_\mathrm{h})&\equiv\frac{f_\mathrm{b}(M_\mathrm{h})}{\Omega_\mathrm{b}/\Omega_\mathrm{m}}\\
    &=\frac{(1+f_\mathrm{blue})\,M_*+M_\mathrm{gas}+f_\mathrm{ICL}\,M_\mathrm{BCG}}{M_\mathrm{h}\left(\Omega_\mathrm{b}/\Omega_\mathrm{m}\right)},
\end{aligned}
\end{equation}
where $f_\mathrm{blue}\sim\mathcal{N}(0.15, 0.17)$ is the fractional mass contribution from blue member galaxies, $f_\mathrm{ICL}\sim\mathcal{N}(0.340, 0.077)$ is the mass contribution from ICL, and $\mathcal{N}(\mu,\sigma)$ annotates a Gaussian distribution with mean $\mu$ and standard deviation $\sigma$. 

The uncertainty of $f_\mathrm{blue}$ and $f_\mathrm{ICL}$ are quoted from~\citep{AEO+22}. Cosmology is marginalized over by drawing $\Omega_\mathrm{m}$ and $\Omega_\mathrm{b}$ from the nine-year WMAP result~\citep{WMAP13}. An additional 15 percent uncertainty is applied to the stellar mass to account for uncertainty in the initial mass function. 
Random realizations of equation~(\ref{eqn:fb_scaling_relation}) are drawn with these uncertainties, and we evaluate the mean and standard deviation of $Y_\mathrm{b}(M_\mathrm{h})$ from the random draw.

We are particularly interested in the mean normalized baryon mass fraction $\bar{Y}_\mathrm{b}$ over halos of mass within $[10^{13},\, 10^{14}]\,M_\odot$ since that is the halo mass range that our $\xi_\pm(\vartheta)$ is most sensitive to, given the redshift distribution of the DES Y3 \textsc{metacal} sample and the scale cut of $\xi_\pm(\vartheta)$~\citep{SMK+23,TPK+24}. 
We calculate the mean and error of $\mathrm{log}\,\bar{Y}_\mathrm{b}$ from the random realizations,
\begin{equation}
\label{eqn:fb_Akino_result}
\begin{aligned}
    \langle \mathrm{log}\,\bar{Y}_\mathrm{b}\rangle &= -0.285,\\
    \sigma\left(\mathrm{log}\,\bar{Y}_\mathrm{b}\right)&=0.076.
\end{aligned}
\end{equation}

Recently, the eROSITA survey~\citep{PBM+24} has found a systematic lower $f_\mathrm{gas}$-$M_\mathrm{h}$ relation than other literature~\citep[e.g.][]{SVD+09,PCA+09,LRS15,EGE+19}. 
\cite{PBM+24} stacks the eROSITA X-ray surface brightness for an optically selected GAMA group sample~\citep{RND+11}, and measures the electron density and gas mass from the stacked X-ray surface brightness.
This avoids the X-ray selection effects, which will bias the $f_\mathrm{gas}$ to higher values if not corrected, since galaxy groups and clusters with low X-ray surface brightness are not included in the sample. The resulting $f_\mathrm{gas}$ measurement is consistent with~\cite{SAM+25}. 
While a detailed discussion of the conflicting gas mass fraction is beyond the scope of this work, we explore the impact of a lower gas mass fraction on our results in Sec.~\ref{sec:6x2cosmo}.

\section{Modeling Baryonic Physics}
\label{sec:baryon}
Since we intend to keep all the cosmic shear data points down to $2.^\prime5$, our analysis requires modeling of baryonic feedback processes. 
Following \citetalias{XEM+23}, we adopt PCA baryon mitigation to account for baryonic effects on small scales~\citep{ekd15,hem19,hem21}. 
In this approach, PCs are constructed directly in the model vector space, which in our case means they span the difference between the gravity-only 2PCFs and the baryonic 2PCFs computed from a range of simulations. After the construction of the PCs, the full model vector is calculated by
\begin{equation}
    \bvec{M}_\mathrm{bary}(\bvec{p},\bvec{Q})=\bvec{M}(\bvec{p})+\sum_{i=1}^{n}Q_i\,\bvec{PC}_i\,,
\end{equation}
where the amplitudes $Q_i$ are varied during analysis, and $\bvec{M}(\bvec{p})$ is the gravity-only model vector. We note that in this paper, we construct PCs for all shear-based auto and cross-correlations. 
This is different from \citetalias{XEM+23}, where it was sufficient to build PCs for the shear auto-correlation 2PCF only, given the larger statistical uncertainty of the DES Y1 dataset. 
For example, a strong feedback scenario in \textsc{ANTILLES} that corresponds to $-2\sigma$ baryon mass fraction can results in $\Delta\chi^2_{w_{sk}}=5.8$ and $\Delta\chi^2_{\gamma_t}=1.5$, which are not negligible for our analysis.

\subsection{Baryon feedback simulations}

In this work, we use the \textsc{ANTILLES} simulations suite~\citep{SMK+23}, a suite comprised of 400 simulations of box size 100 Mpc/h with $256^3$ particles for baryonic and dark matter each. Different simulations have different galaxy formation efficiency (controlled by wind velocity and mass-loading factor of kinetic wind stellar feedback) and AGN feedback strength (controlled by temperature increase, the number of neighboring gas particles to heat, and the gas density threshold to start black hole accretion).
\begin{figure*}
    \centering
    \includegraphics[width=0.8\linewidth]{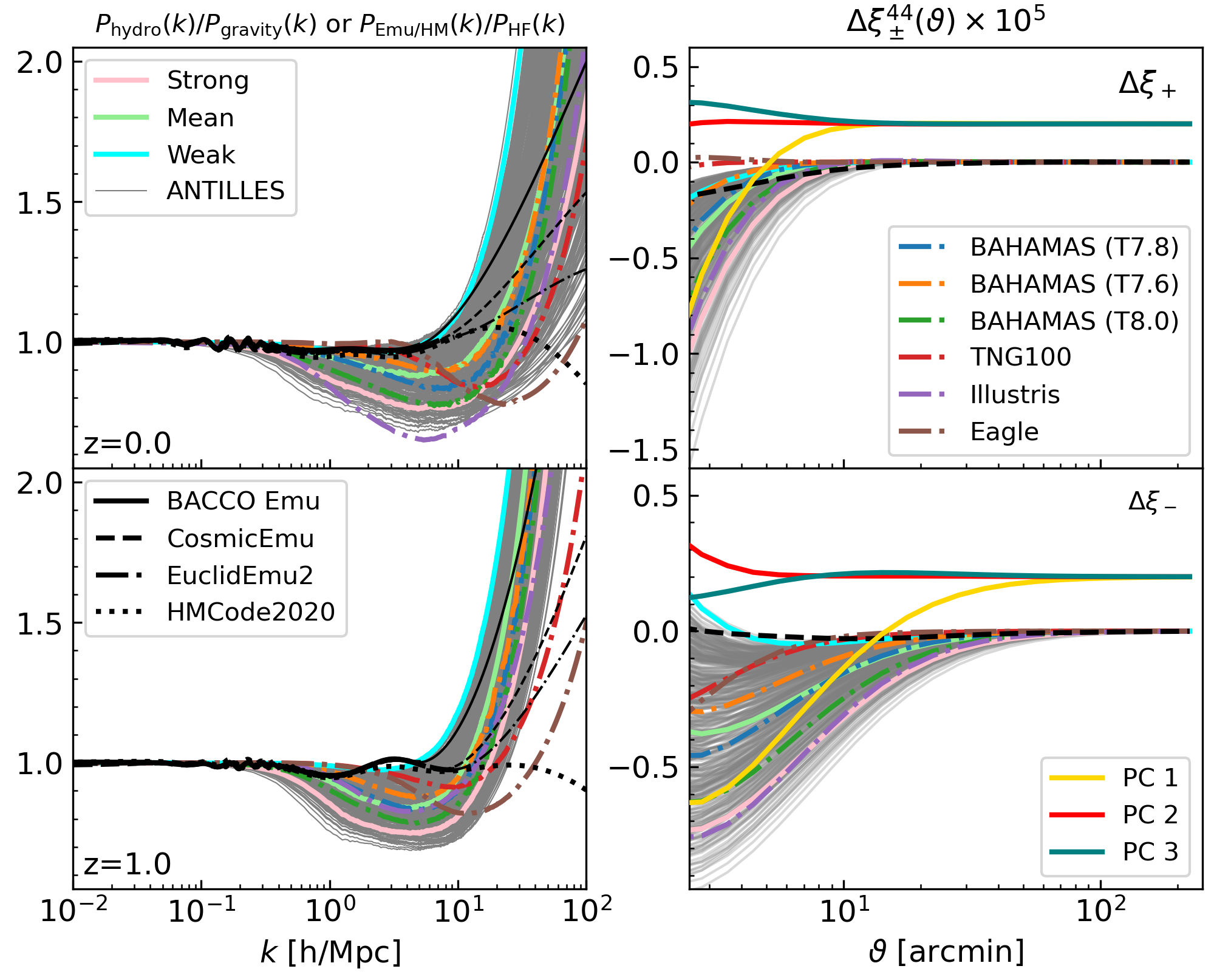}
    \caption{\textbf{Left panels}: matter power spectrum ratio between (i) hydrosims and gravity-only $N$-body sims and (ii) alternative nonlinear power spectrum models and Halofit. 
    The top (bottom) panel shows the ratio at $z=0$ ($z=1$). 
    We show the ratios of ANTILLES simulations in gray solid lines. We show three special ANTILLES simulations in pink, green, and cyan solid lines, whose feedback strength is defined as strong, mean, and weak compared to observation in~\citep{AEO+22} (see Sec.~\ref{sec:fb_Q1}). 
    Other hydrosims are shown in dotted-dashed lines: BAHAMAS $\mathrm{log}\,T_\mathrm{AGN}=7.8$ (blue), $\mathrm{log}\,T_\mathrm{AGN}=7.6$ (orange), $\mathrm{log}\,T_\mathrm{AGN}=8.0$ (green), TNG100 (red), Illustris (purple), and Eagle (brown). 
    Black lines show the contrast between Halofit~\citep{SPJ+03,TH13,BMSP14} and alternative nonlinear $P(k)$ models: \textsc{BACCOemu} (solid), \textsc{CosmicEmu} (dashed), \textsc{EuclidEmu2} (dotted-dashed), and \textsc{HMCode2020} (dotted). The thick black lines show the $P(k)$ ratios in the native $k$ range of the alternative models and the thin black lines show their spline extrapolation. 
    \textbf{Right panels}: $\xi_\pm^{44}(\vartheta)$ difference due to baryonic feedback and nonlinear power spectrum modeling. We show the first three PCs in solid yellow/red/teal lines, which are padded along the $y$-axis for better visualization. 
    Hydrosims are shown with the same colors and line styles as the left panels. 
    We only show one alternative $P(k)$ model, \textsc{EuclidEmu2}, in the left panels.}
    \label{fig:Pk_antilles}
\end{figure*}

\begin{figure}
    \centering  \includegraphics[width=\linewidth]{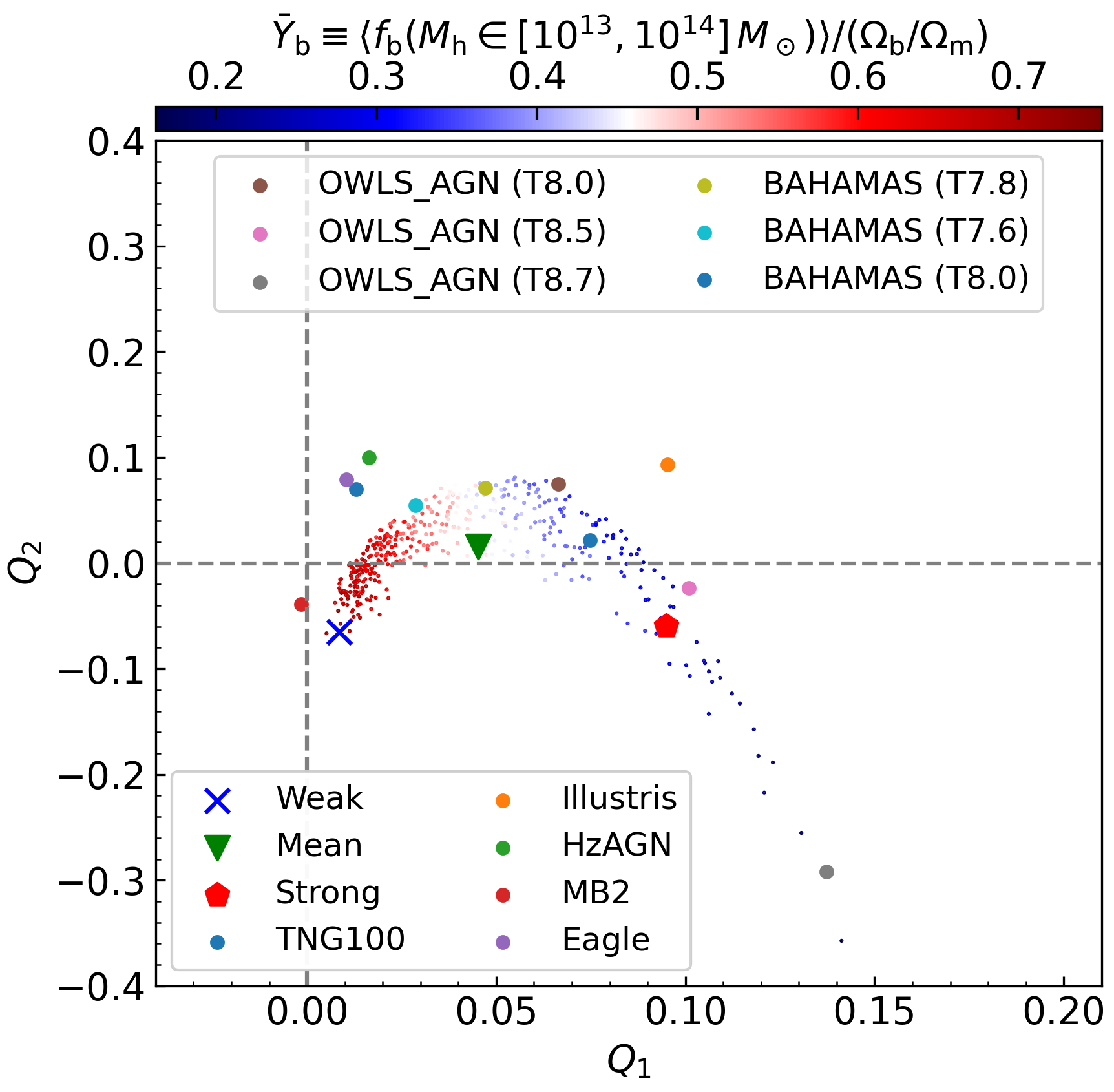}
    \caption{The amplitudes of the first two PCs ($Q_1$ and $Q_2$) of the hydrodynamical simulations used in this work. The smallest data points are \textsc{ANTILLES} simulations that are used to generate the PCs, and are colored based on their mean normalized baryon fraction $\bar{Y}_\mathrm{b}$ of halos within $10^{13}$-$10^{14}$ $M_\odot$. Note that $\bar{Y}_\mathrm{b}$ correlates with $Q_1$ tightly. The medium data points are the 11 hydrosims used in~\citetalias{XEM+23}. The largest data points are the three \textsc{ANTILLES} simulations that are reserved from PC generation, corresponding to the weak (blue cross), mean (green triangle), and strong (red pentagon) feedback scenarios.}
    \label{fig:Qs_antilles}
\end{figure}

\begin{figure}
    \centering
    \includegraphics[width=\linewidth]{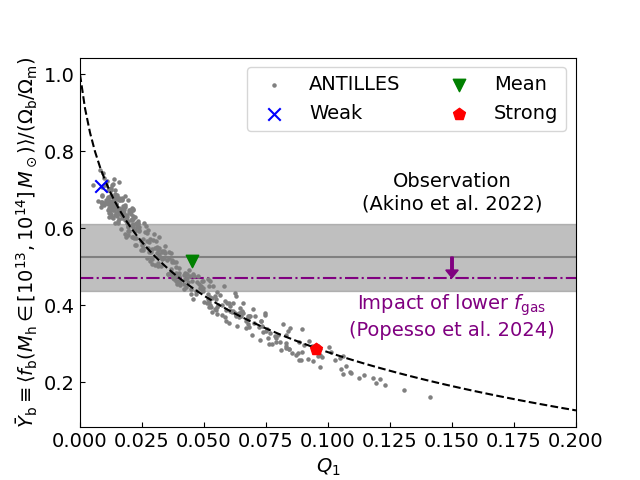}
    \caption{The scaling relation between $Q_1$ and the mean baryon fraction in halos of mass $M_\mathrm{h}\in[10^{13},\,10^{14}]\,M_\odot$, normalized by the cosmic baryon fraction $\Omega_\mathrm{b}/\Omega_\mathrm{m}$. The gray data points are \textsc{ANTILLES} simulations used to generate PCs, and the colored data points are the weak (blue cross), mean (green triangle), and strong (red pentagon) feedback scenarios. 
    The relation between $Q_1$ and $\bar{Y}_\mathrm{b}$ can be fitted as the black dashed line.
    We show the $\bar{Y}_b$ observed in \cite{AEO+22} in the horizontal solid line and its uncertainty in the shaded gray region. Alternative estimate of $\bar{Y}_b$ based on the latest eROSITA $f_\mathrm{gas}$ measurement~\citep{PBM+24} is shown in the horizontal purple line. We note that this is not a rigorous estimate since it assumes that the HSC-XXL sample and the eROSITA sample have the same stellar mass fraction, and we only aim to show the impact qualitatively.}
    \label{fig:Q1_fb}
\end{figure}

The 400 simulations are carefully chosen to uniformly sample the feedback parameters using 200 Latin hypercube nodes, and each node is run with two hydrodynamic solvers (smoothed-particle hydrodynamics or classic SPH, and Anarchy SPH). The simulations have been demonstrated to be sufficiently large for capturing relative effects of baryonic to dark matter power spectrum \citep{MSB+17,MBS+18,DMS20} and span a large range of feedback, including extreme feedback scenarios. 

To test the PCA baryonic feedback mitigation, we select three special feedback scenarios from the ANTILLES simulations and exclude them during PC generation. The three scenarios are chosen based on their mean baryon mass fraction within halo mass range $[10^{13},\, 10^{14}]\,M_\odot$. They are consistent, $2\sigma$ higher, or $2\sigma$ lower than the observed baryon mass fraction in ~\citep{AEO+22}, and we refer to them as the mean, weak, and strong feedback, since stronger feedback leads to lower baryon mass fraction.

We show the range of small-scale $P(k)$ suppression spanned by ANTILLES simulations (pink, green, and blue solid lines for three special \textsc{ANTILLES} scenarios, and grey lines for the other \textsc{ANTILLES} simulations) and other simulations used in~\citetalias{XEM+23} in the left panels of Fig.~\ref{fig:Pk_antilles}. 

These figures also show the impact of different models of the nonlinear matter power spectrum (\textsc{BACCOemu}~\citep{AAC+21,AZC+21}, \textsc{CosmicEmu}~\citep{HBL+16,LHK+17}, \textsc{EuclidEmu2}~\citep{EuclidEmu2}, and \textsc{HMCode2020}~\citep{MBT+21}) in the left panels of Fig.~\ref{fig:Pk_antilles} as a comparison. 
Note that the uncertainty from nonlinear matter power spectrum can be comparable to baryonic feedback at $k\geq10\,h/\mathrm{Mpc}$. We explore possible cross-talk between the two systematics in Sec.~\ref{sec:systematics}.

In the right panels of Fig.~\ref{fig:Pk_antilles}, we show the corresponding bias in $\xi_\pm^{44}(\vartheta)$ due to different feedback scenarios and nonlinear matter power spectrum models. We also show the first three PCs constructed from the remaining 397 \textsc{ANTILLES} simulations. 

We derive the PC amplitudes $Q_i$ for \textsc{ANTILLES} and other simulations used in~\citep{hem21} and show the distribution of $Q_1$-$Q_2$ in Fig.~\ref{fig:Qs_antilles}. 
We note that the value of $Q_i$ is survey-dependent since PCs are calculated from precision matrix-weighted data vectors. Therefore, $Q_i$ in this work can not be compared to the ones in~\citetalias{XEM+23} directly. 
We define the wide prior of $Q_1$ as a uniform distribution between $[-20, 20]$, and explore a more informative, baryon mass fraction-based prior in Sec.~\ref{sec:fb_Q1}.

\subsection{Calibrating baryonic feedback from baryon mass fraction observations}
\label{sec:fb_Q1}

Several publications~\citep{DMS20,DAT+23,PLB+23,SMK+23,LD24,DKB+25} have demonstrated that small-scale suppression of the matter power spectrum can be predicted from the baryon mass fraction. 
In order to include corresponding measurements of the baryon mass fraction in our analyses, we first derive a relation between the mean normalized baryon mass fraction within $10^{13}$ to $10^{14}\, M_\odot$ halos and our PCA amplitude parameters $Q_i$.

The presence of this relation in the \textsc{ANTILLES} simulations was already confirmed in \citep{SMK+23}. Its physical origin stems from the gravitational nature of baryonic feedback in redistributing matter around dark matter halos~\citep{DMS20}.
\cite{SMK+23} measures the baryon mass fraction for each simulation in the \textsc{ANTILLES} suite. The measurement accounts for all stellar and gas particles within a spherical overdensity radius. 

We show the $Q_1$ and $Q_2$ of the \antilles\,simulation and the 11 hydrodynamical simulations used in~\citetalias{XEM+23} in Fig.~\ref{fig:Qs_antilles}. 
We also color-coded the \antilles\,simulation data points with the mean normalized baryon mass fraction $\bar{Y}_\mathrm{b}$ between $10^{13}$ and $10^{14}\,M_\odot$ halos. 
The prominent color gradient indicates a tight correlation between $Q_1$ and $\bar{Y}_\mathrm{b}$. 
Fig.~\ref{fig:Q1_fb} directly compares $Q_1$ and $\bar{Y}_\mathrm{b}$, and we find a nonlinear correlation with low dispersion. We also show the prediction from~\citep{AEO+22} as the horizontal gray band.
We design a parametric fitting function $\mathcal{F}(Q_1)$ for $\bar{Y}_\mathrm{b}$ and $Q_1$,
\begin{equation}
\label{eqn:Yb_Q1_fit}
    \bar{Y}_\mathrm{b}\equiv \mathcal{F}(Q_1) = \frac{\mathrm{ln}((Q_b+Q_1)/Q_p)}{\mathrm{ln}(Q_b/Q_p)},
\end{equation}
where $Q_b$ is the pivot $Q_1$ and $Q_p$ is the characteristic scale of $Q_1$. 
The functional form is chosen such that $\bar{Y}_\mathrm{b} \equiv 1$ when $Q_1=0$, i.e., no feedback. 
The best-fitting coefficients are $Q_b=0.0042$ and $Q_p=0.36$, and we show the corresponding relation as the black dashed line in Fig.~\ref{fig:Q1_fb}. 

We assume that $\bar{f}_\mathrm{b}/(\Omega_\mathrm{b}/\Omega_\mathrm{m})$ measured in~\cite{AEO+22} follows a lognormal distribution. Therefore, we can introduce a prior for $Q_1$ based on the baryon mass fraction observation
\begin{equation}
\label{eqn:Q1_fb_prior}
    \mathcal{\pi}(Q_1)=-\frac{1}{2}\left(\frac{\mathrm{log}\mathcal{F}(Q_1) -  (-0.285)}{0.082}\right)^2,
\end{equation}
where $\mathcal{\pi}(Q_1)$ is the log prior and $-0.285$ is the mean $\mathrm{log}\,\bar{Y}_\mathrm{b}$ measured in~\cite{AEO+22}. The scatter $0.082$ here also includes the $0.029$ dex intrinsic scatter of the $\bar{Y}_\mathrm{b}$-$Q_1$ relation in Fig.~\ref{fig:Q1_fb}.

We note that although we do not have the baryon mass fraction measurement of the other 11 hydrodynamical simulations, a similar relation between baryon mass fraction and matter power spectrum suppression has been found in \cite{DMS20}. At a fixed scale, $P(k)$ suppression is proportional to $Q_1$, and we expect other simulations to follow a similar $\bar{Y}_\mathrm{b}$-$Q_1$ trend. Given that both the PCs and $\bar{Y}_\mathrm{b}$ are measured from \antilles, our method is self-contained, and exploring this method on a larger heterogeneous simulation library is beyond the scope of this paper.

\section{Machine Learning Accelerated Inference}
\label{sec:inference}
Due to the increased constraining power of the planned analysis, we conduct an extensive study of the impact of a variety of systematics (and their cross-talks). In order to complete these studies in finite time, we train a neural network emulator to evaluate the $6\times2$pt model vector as a function of cosmological and nuisance parameters. 
This strategy has been developed in~\citep{BEM+23,BEM+24,ZSM+23,TRK+23,SZM+24,AHE24,SKD+25,ZSK+25}. Below we detail calculations of model vector and covariances and describe the setup and accuracy of our neural network emulator.

\subsection{Theory models, covariances - Cobaya-CosmoLike Architecture}

We adopt the \textsc{CoCoA}~\citep{cocoa,EKS+14,KE17,TL21,FES+22} inference pipeline for likelihood analyses and validate our pipeline against the public DES Y3 $3\times2$ analysis and Planck PR4 CMB lensing analysis~\citep[][also see Appendix~\ref{sec:code_compare}]{DES_Y3_3x2pt,PCF+21,CML22}.

\subsubsection{Data vector modeling}
We calculate the angular bin-averaged results for the five position-space correlation functions as: 
\begin{subequations}\label{eqn:dv_theory}
    \begin{align}
        \xi_\pm^{ab}(\vartheta) &= \sum_\ell \frac{2\ell+1}{2\pi\ell^2(\ell+1)^2}\left(\overline{G_{\ell,2}^+\pm G_{\ell,2}^-}\right)\,C_{\kappa_g\kappa_g}^{ab}(\ell)\,, \label{eqn:dv_xipm}\\
        \gamma_{t}^{ab}(\vartheta) &= \sum_\ell \frac{2\ell+1}{4\pi\ell(\ell+1)}\overline{P_\ell^2}\,C_{\delta_g\kappa_g}^{ab}(\ell)\,, \label{eqn:dv_gammat}\\
        w_{g}^{a}(\vartheta) &= \sum_\ell \frac{2\ell+1}{4\pi}\overline{P_\ell}\,C_{\delta_g\delta_g}^{aa}(\ell)\,, \label{eqn:dv_wtheta}\\
        w_{g\kappa}^a(\vartheta) &= \sum_\ell \frac{2\ell+1}{4\pi}F(\ell)\,\overline{P_\ell}\,C_{\delta_g\kappa}^{a}(\ell)\,, \label{eqn:dv_w_gk}\\
        w_{s\kappa}^a(\vartheta) &= \sum_\ell \frac{2\ell+1}{4\pi\ell(\ell+1)}F(\ell)\,\overline{P_\ell^2}\,C_{\kappa_g\kappa}^{a}(\ell)\,, \label{eqn:dv_w_sk}
    \end{align}
\end{subequations}
where $\overline{G_{\ell,2}^+\pm G_{\ell,2}^-}$, $\overline{P_\ell}$, and $\overline{P_\ell^2}$ are $\theta$-bin-averaged Legendre polynomials defined in Ref.~\citep{FKEM20}. The exact mathematical description of the corresponding power spectra $C_{XY}^{ab}(\ell)$ can be found in \citetalias{XEM+23}.

$F(\ell)$ is a window function incorporating the Gaussian smoothing, pixelization, and scale cut that we apply to the CMB lensing convergence map when measuring the cross-correlations,
\begin{equation}
\begin{aligned}
    \label{eqn:beam_kernel}
    F(\ell) =&\widetilde{W}_{N_\mathrm{side}}(\ell)\Theta(\ell-L_\mathrm{min})\Theta(L_\mathrm{max}-\ell)\\
    &\times\mathrm{exp}(-\ell(\ell+1)/\ell_\mathrm{beam}^2)\,,
\end{aligned}
\end{equation}
where $\Theta(\ell)$ is the Heaviside function and $\ell_\mathrm{beam}\equiv\sqrt{16\mathrm{ln}2}/\theta_\mathrm{FWHM}$. We assume $\theta_\mathrm{FWHM}=7\arcmin$ in this work. 
$(L_\mathrm{min},L_\mathrm{max})=(8,2048)$ are the scale cuts in the CMB lensing convergence map. $\widetilde{W}_{N_\mathrm{side}}(\ell)$ is the pixel window function of \textsc{HEALPix}. 

Since the auto-correlation of CMB lensing enters the data vector as a Fourier space quantity, the corresponding model vector $C_{L_{b}}^{\kappa\kappa}$ is calculated as
\begin{equation}
    \label{eqn:dv_bp_marg}
    \begin{aligned}
    C_{L_{b}}^{\kappa\kappa} =& \sum_L\left(\mathcal{B}_i^L+M_i^{\phi,L}\right)C_L^{\kappa\kappa}\ - \Delta C_{L_{b}}^{\kappa\kappa,b}\,,
    \end{aligned}
\end{equation}
where $\Delta C_{L_{b}}^{\kappa\kappa,b}$ is a constant bias pre-computed at the FFP10 fiducial cosmology
\begin{equation}
\begin{aligned}
    \Delta C_{L_{b}}^{\kappa\kappa,b} \equiv& \sum_L  \frac{L^2(L+1)^2}{4}M_i^{X,\ell}\left(\left.C_\ell^X\right|_\mathrm{fid}-\hat{C}_\ell^{X}\right)\\
    &+\sum_L M_i^{\phi,L}\left.C_L^{\kappa\kappa}\right|_{\mathrm{fid}}\,,
\end{aligned}
\end{equation}
here $X$ sums over the eight MV lensing estimators (TT, EE, TE, TB, EB, ET, BT, and BE). 
$\mathcal{B}_i^L$, $M_i^{\phi,L}$, and $M_i^{X,L}$ are binning and correction matrices which can be found in~\citetalias{XEM+23} and \cite{P18A8}, $C_L^{\kappa\kappa}$ is theoretical convergence power spectrum at sampled cosmology, $C_\ell^{X}|_\mathrm{fid}$ is CMB lensing and primary CMB power spectra precomputed at FFP10 cosmology, and $\hat{C}_\ell^X$ is the estimated primary CMB power spectra.

\subsubsection{Covariance modeling}
The covariance matrix of the $6\times2$pt likelihood is evaluated using \textsc{CosmoCov}~\citep{KE17,FKEM20,FAC+21,FES+22} following~\citetalias{XEM+23} and includes corrections from survey geometry. 
The covariance matrix is analytical except for the CMB lensing auto-covariance, for which we adopt the numerical covariance from~\cite{CML22} to marginalize over uncertainties from the primary CMB power spectra and other intricacies in CMB lensing reconstruction. Uncertainties related to survey systematics in galaxy clustering measurements~\citep{RWE+22}, point mass marginalization~\citep{MBJ+20,DESY3_method,PKD+22} are also accounted for in the covariance. 

The covariance matrix is generated at a fiducial $\Lambda$CDM with $(\Omega_\mathrm{m},\,\sigma_8,\,n_s,\,\Omega_\mathrm{b},\,H_0)=(0.318,\,0.744,\,0.871,\,0.040,\,0.678)$. We adopt the survey area $\Omega_\mathrm{s}=4143.2$ deg$^2$, and a single-component shape noise of $\sigma_\epsilon=(0.243,\,0.262,\,0.259,\,0.301)$ for each tomography bin. 
When calculating the covariance, we fix linear galaxy bias to $(1.448,\,1.666,\,1.860,\,1.756,\,1.924,\,1.865)$ and lens galaxy magnification bias to $(0.430,\,0.300,\,1.750,\,1.940,\,1.560,\,2.960)$. 
We calculate the covariance matrix in curved-sky geometry, assuming the Limber approximation. 
When evaluating the covariance matrix blocks related with $C^{\kappa\kappa}_{L_\mathrm{b}}$, we sum over $L\in[8,\,2500]$. 

\subsection{Neural network details and performance}
\label{sec:emu}

\subsubsection{Architecture}
\label{sec:emu_arch}
\begin{figure}
    \centering
    \includegraphics[width=0.5\linewidth]{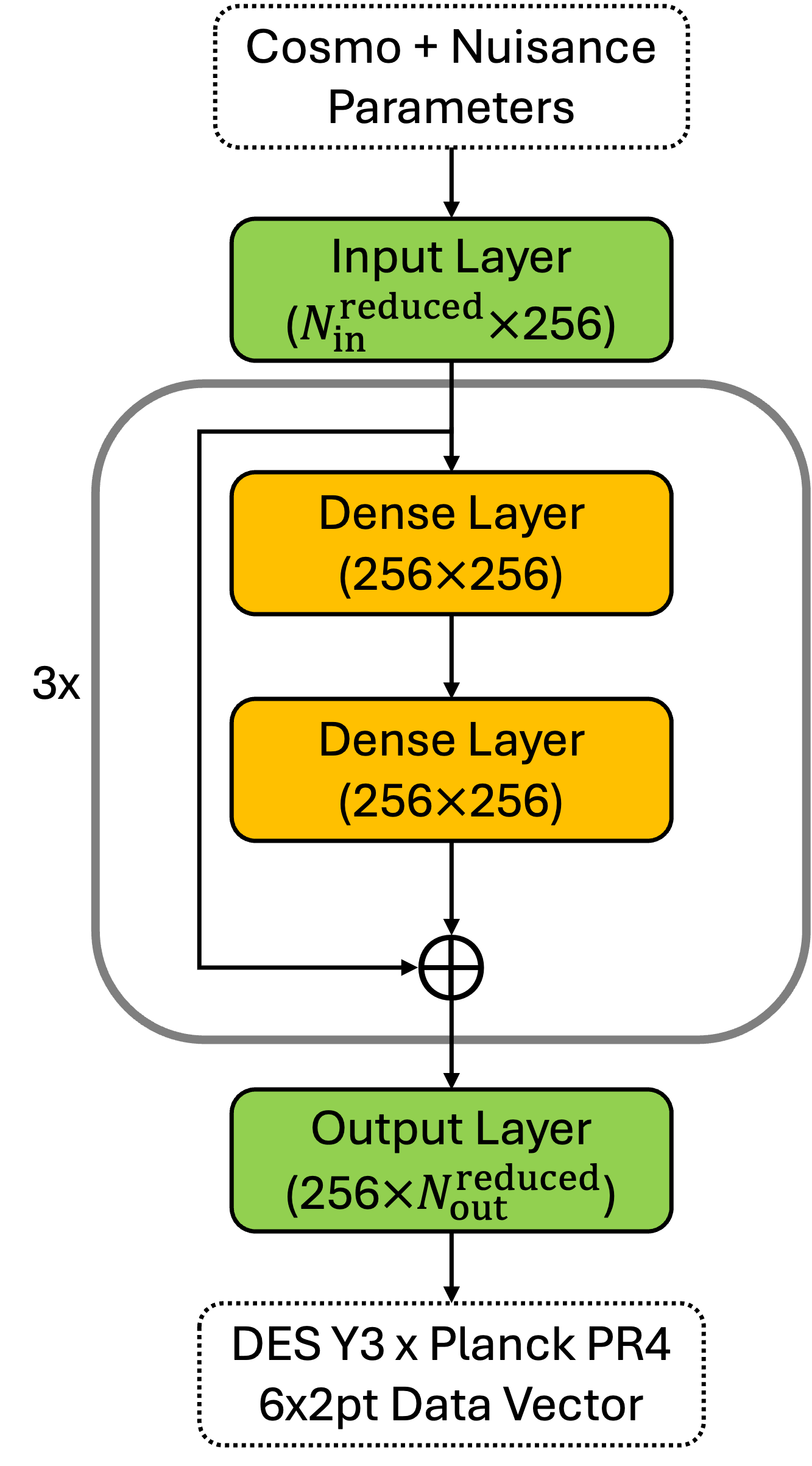}
    \caption{The architecture of the neural network emulator for a single probe. The input cosmological and nuisance parameters are masked and normalized such that only relevant parameters are passed into the input layer of size $N_\mathrm{in}^\mathrm{reduced}\times256$, which applies a linear affine transform to the input data. Three $256\times256$ residual networks (\textsc{ResMLP}) layers are appended to the input layer, each consisting of two dense layers with a residual connection. Then the $256\times N_\mathrm{out}^\mathrm{reduced}$ output layer is appended to the \textsc{ResMLP} layers to predict a specific probe in $6\times2$pt.}
    \label{fig:NN_arch}
\end{figure}

We adopt a similar neural network architecture as developed in~\citep{SZM+24} (see Fig.~\ref{fig:NN_arch}). We train one of these neural networks for each of the six two-point statistics in our data vector \{$\xi_\pm^{ab}(\vartheta)$, $\gamma_t^{ab}(\vartheta)$, $w_{gg}^{a}(\vartheta)$, $w_{g\kappa}^a(\vartheta)$, $w_{s\kappa}^a(\vartheta)$, $C_{L_\mathrm{b}}^{\kappa\kappa}$\} independently. 
We separate the input parameters into fast and slow parameters. Fast parameters, including shear calibration bias and baryonic feedback PC amplitude, do not require complex calculations. Their impact can be instantly applied to the model vector.
Slow parameters, including cosmological parameters, photo-$z$ bias and stretch, linear galaxy bias, IA amplitude and redshift evolution, involve complex calculations. Therefore, we only train emulators in the slow parameter space and apply the fast parameters to the emulator prediction on the fly.

For each emulator, cosmological and nuisance parameters are masked, normalized, and fed into an input affine linear layer of $N_\mathrm{in}^\mathrm{reduced}\times256$, where $N_\mathrm{in}^\mathrm{reduced}$ is the number of unmasked input parameters that are relevant to this probe. 
For example, the $\xi_\pm^{ab}(\vartheta)$ emulator has $N_\mathrm{in}^\mathrm{reduced}=12$ including 6 cosmological parameters, 4 photo-$z$ bias parameters, and 2 NLA parameters. The input layer is followed by three $256\times256$ multi-layer perceptron (MLP) with residual connection (\textsc{ResMLP}), and then a linear output layer of $256\times N_\mathrm{out}^\mathrm{reduced}$. 
Predictions from the emulators are then post-processed and concatenated to return the full model vector for any given analysis. 

\subsubsection{Training}
\label{sec:emu_train}
We follow~\citep{SZM+24} in generating the training and validation samples. We first run a synthetic $6\times2$pt analysis on mock data and calculate the parameter covariance. 
We temper the parameter covariance $\bvec{C}$ by a temperature $T=128$\footnote{We abbreviate the emulator neural network model trained by $T=128$ data as \textsc{ResMLP128} hereafter.} such that $\bvec{C}\mapsto T\bvec{C}$, and generate 1 million training data by sampling the tempered covariance. The validation data, which contains 10,000 samples, is generated from the same covariance but with a temperature of $T=64$. We remove data points with $\Delta\chi^2>10^{5}$ compared to the fiducial data vector to stabilize the training. 
We note that $T=128$ is a temperature large enough such that the emulator is accurate over the posterior of $\xi_\pm(\vartheta)$, $3\times2$pt, c$3\times 2$pt, and $6\times2$pt analyses. 

During training, the input cosmological and nuisance parameters are first masked and standardized, ensuring that only relevant parameters are passed to the input layer, and the parameters have a zero mean and unit standard deviation.
The data vectors $\bm{y}$ are also standardized to $\tilde{\bm{y}}$ as
\begin{equation}
    \tilde{\bm{y}}=\frac{\bvec{U}^{-1}\cdot(\bm{y}-\bm{y}_\mathrm{fid})}{(\bvec{U}\cdot\bvec{C}\cdot\bvec{U}^{-1})^{1/2}},
\end{equation}
where $\bvec{U}$ is the eigenbasis of the covariance matrix $\bvec{C}$ and $\bm{y}_\mathrm{fid}$ is the fiducial data vector. 

We use generalized Swish activation function~\citep{swish} for the \textsc{ResMLP} layers
\begin{equation}
    f(x)=\left(\gamma  + \frac{1-\gamma}{1+e^{-\beta x}}\right)x,
\end{equation}
where $\beta$ and $\gamma$ are trainable parameters. 

The loss function $L(\bm{w})$ is a hyperbola of $\Delta\chi^2$ between the training data produced by \textsc{CoCoA} $\bm{y}_\mathrm{CoCoA}$ and the neural network prediction $\bm{y}_\mathrm{NN}$
\begin{equation}
    \begin{aligned}
        \Delta\chi^2 &\equiv (\bm{y}_\mathrm{CoCoA}-\bm{y}_\mathrm{NN})^\mathrm{T}\cdot\bvec{C}^{-1}\cdot(\bm{y}_\mathrm{CoCoA}-\bm{y}_\mathrm{NN}),\\
        L(\bm{w}) &= \langle \sqrt{1+2\Delta\chi^2}\rangle-1,
    \end{aligned}
\end{equation}
where $\bm{w}$ is a vector of all trainable parameters, and $\langle\,\rangle$ means sample average. We use the \textsc{Adam} optimizer~\citep{ADAM} with an adaptive learning rate to train the network. We use the $L^2$-norm of the model weights to penalize overfitting, and the $L^2$ penalty is controlled by a weight decay of $10^{-3}$.  
We choose a batch size of 256 and train 200 epochs for each emulator. 
The learning rate starts from $10^{-3}$ and is reduced by a factor of 10 when validation loss plateaus for 10 epochs.

\subsubsection{Validation}
\label{sec:emu_valid}
\begin{figure}
    \centering
    \includegraphics[width=\linewidth]{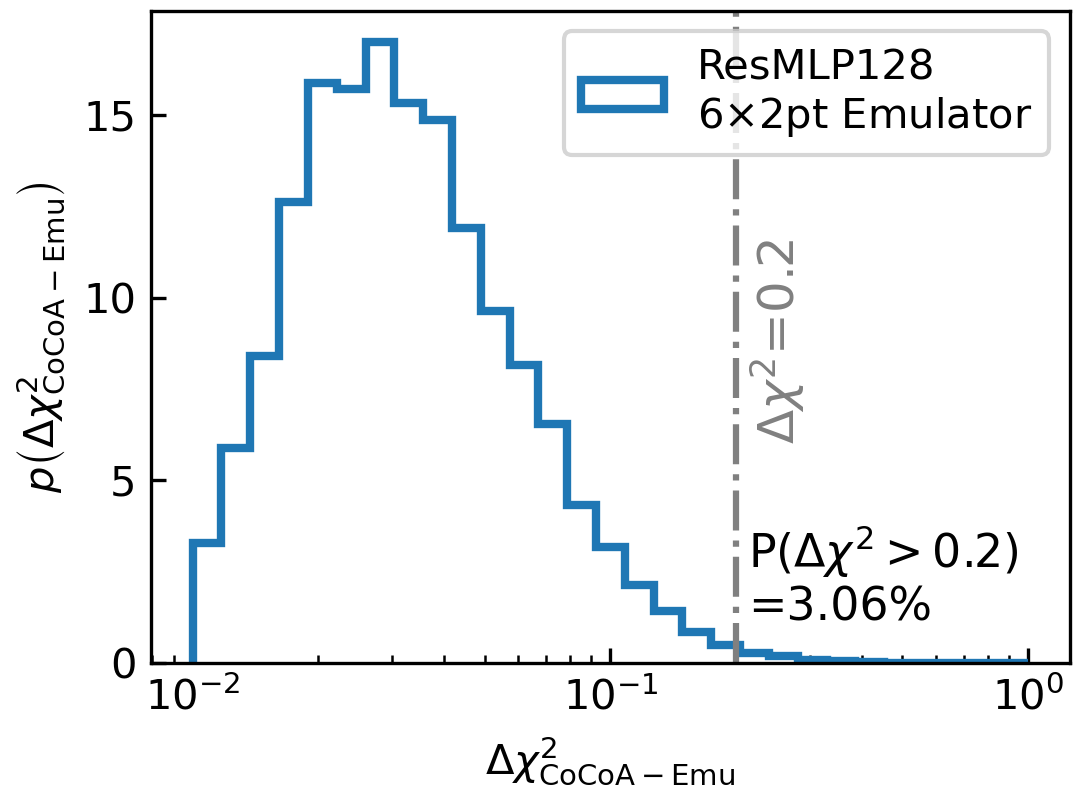}
    \caption{The distribution of $\Delta\chi^2$ between the $6\times2$pt model vectors evaluated by \textsc{CoCoA} and the \textsc{ResMLP128} emulator over the validation dataset, which has $T=64$. Over the 10,000 samples in the validation dataset, only 3.06 percent of the samples give $\Delta\chi^2>0.2$. }
    \label{fig:emu_valid_dchi2}
\end{figure}

\begin{figure}
    \centering
    \includegraphics[width=\linewidth]{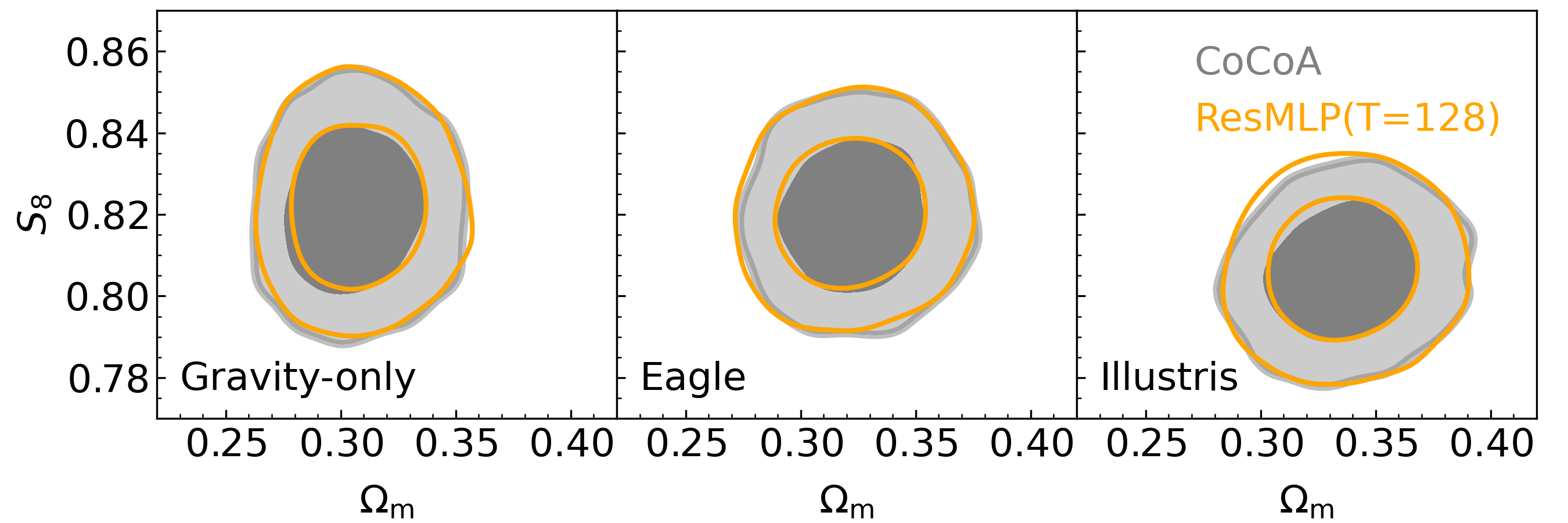}
    \caption{Posterior distribution in $\Omega_\mathrm{m}$ and $S_8$ for synthetic analyses from \textsc{CoCoA} (filled contours) and the \textsc{ResMLP128} emulator (unfilled contours). The fiducial data vectors are gravity-only (left panel), contaminated by Eagle-like feedback (middle panel), and Illustris-like feedback (right panel). 
    All analyses include cosmic shear data points down to $0.^\prime25$ and marginalize over two PCs with wide priors to test the accuracy of the emulator.}
    \label{fig:emu_valid_post}
\end{figure}
Our first validation metric is to evaluate the $\Delta\chi^2$ between the $6\times2$pt data vectors predicted by \textsc{CoCoA} and \textsc{ResMLP128} over the validation sample. The result is shown in Fig.~\ref{fig:emu_valid_dchi2}. We only see a 3.06 percent probability for $\Delta\chi^2>0.2$, which is the theoretical error threshold adopted by DES Y3 without importance sampling correction. 

We also compare posteriors sampled by MCMC chains run by \textsc{CoCoA} and \textsc{ResMLP128} using the fiducial cosmology with three baryonic feedback scenarios as input: gravity only, Eagle, and Illustris.  
We include all the data points in $\xi_\pm(\vartheta)$ down to 0$\farcm$25 and adopt the baseline scale cut for all the other probes. Posteriors are shown in Fig.~\ref{fig:emu_valid_post} and show excellent agreement of contours when using the emulator versus the brute force calculation.

We conclude that our emulator is reliable for the subsequent impact studies, mock analyses, and analysis choice decisions.

\section{Analysis Choices and Simulated Analyses}
\label{sec:choices}

\subsection{Systematics models}
\label{sec:systematics}
We follow~\citep{XEM+23,DES_Y3_3x2pt} for systematics modeling in shear calibration bias, photo-$z$ uncertainty, linear galaxy bias, point-mass marginalization, and magnification bias. 
The baryonic feedback mitigation method, including external prior from baryon mass fraction observation, is discussed in Sec.~\ref{sec:baryon}.
Thus, we only discuss the remaining systematic modeling that differs from the DES Y3 Key Project~\citep{DES_Y3_3x2pt} in this section, including IA models (Sec.~\ref{sec:sys_IA}), point-spread function (PSF) uncertainties~\ref{sec:sys_psf}, and nonlinear matter power spectrum modeling (Sec.~\ref{sec:sys_nl}).
Our baseline analysis settings, including parameter spaces and corresponding priors, are summarized in Table~\ref{tab:params_table}.

\begin{table}[h]
	\centering
	\begin{tabular}{lcc}
	\toprule
    \rule{0pt}{5ex}
	Parameters & \shortstack{Fid. Value} & Prior\\
	\tableline
    \multicolumn{3}{l}{\texttt{Cosmological Parameters}}\\
	$\Omega_\mathrm{m}$& 0.3 & flat[0.1, 0.9]\\
	$ A_s\times 10^9 $& 2.1 & flat[0.5, 5.0]\\
	$n_s$& 0.96605 & flat[0.87, 1.07]\\
	$\Omega_\mathrm{b}$& 0.040 & flat[0.03, 0.07]\\
	$H_0$& 67.32 & flat[55, 91]\\
    $\sum m_\nu$/eV & 0.06 & flat[0.0558, 0.12]\\
	\multicolumn{3}{l}{\texttt{Source Photo-$z$}}\\
	$\Delta_{z,\,\mathrm{src}}^1\times 10^2$ & 0.0 & $\mathcal{N}(0.0, 1.8)$ \\
	$\Delta_{z,\,\mathrm{src}}^2\times 10^2$ & 0.0 & $\mathcal{N}(0.0, 1.5)$ \\
	$\Delta_{z,\,\mathrm{src}}^3\times 10^2$ & 0.0 & $\mathcal{N}(0.0, 1.1)$ \\
	$\Delta_{z,\,\mathrm{src}}^4\times 10^2$ & 0.0 & $\mathcal{N}(0.0, 1.7)$ \\
	\multicolumn{3}{l}{\texttt{Maglim Lens Photo-$z$}}\\
	$\Delta_{z,\,\mathrm{lens}}^1\times 10^2$ & -0.9 & $\mathcal{N}(-0.9, 0.7)$ \\
	$\Delta_{z,\,\mathrm{lens}}^2\times 10^2$ & -3.5 & $\mathcal{N}(-3.5, 1.1)$ \\
	$\Delta_{z,\,\mathrm{lens}}^3\times 10^2$ & -0.5 & $\mathcal{N}(-0.5, 0.6)$ \\
	$\Delta_{z,\,\mathrm{lens}}^4\times 10^2$ & -0.7 & $\mathcal{N}(-0.7, 0.6)$ \\
    $\sigma z^1_\mathrm{lens}$ & 0.975 & $\mathcal{N}(0.975, 0.06)$ \\
    $\sigma z^2_\mathrm{lens}$ & 1.306 & $\mathcal{N}(1.306, 0.09)$ \\
    $\sigma z^3_\mathrm{lens}$ & 0.870 & $\mathcal{N}(0.870, 0.05)$ \\
    $\sigma z^4_\mathrm{lens}$ & 0.918 & $\mathcal{N}(0.918, 0.05)$ \\
	\multicolumn{3}{l}{\texttt{Shear Calibration Bias}}\\
	$m^1\times 10^2$ & -0.63 &  $\mathcal{N}(-0.63, 0.9)$ \\
    $m^2\times 10^2$ & -1.98 &  $\mathcal{N}(-1.98, 0.8)$ \\
    $m^3\times 10^2$ & -2.41 &  $\mathcal{N}(-2.41, 0.8)$ \\
    $m^4\times 10^2$ & -3.69 &  $\mathcal{N}(-3.69, 0.8)$ \\
	\multicolumn{3}{l}{\texttt{Baryonic Feedback}}\\
    \multirow{2}{*}{$Q_1$} & \multirow{3}{*}{\shortstack[l]{gravity/Eagle/Illustris/\\strong/mean/weak}} & wide: flat[-20, 20]\\ 
    & & $\bar{Y}_\mathrm{b}$-based: Eqn.~(\ref{eqn:Q1_fb_prior})\\
	$Q_2$ & & flat[-20, 20]\\
	\multicolumn{3}{l}{\texttt{Intrinsic Alignment}}\\
	$A_\mathrm{IA}$ & -0.7 & flat[-5, 5]\\
	$\beta_\mathrm{IA}$ & -1.7 & flat[-5, 5]\\
	\multicolumn{3}{l}{\texttt{Maglim Linear Galaxy Bias}}\\
	$b_g^{1...4}$ & (1.5, 1.8, 1.8, 1.9) & flat[0.8, 3.0]\\
     \multicolumn{3}{l}{\texttt{Maglim Magnification Bias}}\\
    $C^{1...4}$ & (0.43, 0.30, 1.75, 1.94) & fixed \\
     \tableline
	\end{tabular}
	\caption{This table summarizes the prior ranges used in both the synthetic and real analyses and the fiducial parameter values used in synthetic analyses. We express a flat prior as flat[min, max] and a Gaussian prior as $\mathcal{N}(\mu,\sigma)$ where $\mathcal{N}$ means normal distribution, $\mu$ is the mean, and $\sigma$ is the standard deviation. 
    We choose a narrow prior for neutrino mass to avoid projection effect. 
    Six fiducial baryonic feedback scenarios are adopted in our synthetic analyses: gravity-only, Eagle, Illustris, \antilles-1 (strong), \antilles-100 (mean), and \antilles-379 (weak). $Q_2$ is fixed in our baseline analysis and is sampled with a flat prior $[-20,\,20]$ during the impact study. 
    We note that we only use the first four tomography bins of the \textsc{Maglim} sample, so we omit the nuisance parameters of the last two bins.}
	\label{tab:params_table}
\end{table}

\subsubsection{Intrinsic alignment (IA)}
\label{sec:sys_IA}
Recent work~\citep{FHJ+21} found that large-scale IA is mostly due to the alignment of red central galaxies. At small scales, the IA contamination is heavily influenced by satellite galaxy alignment in the 1-halo regime, and this satellite alignment becomes increasingly significant towards lower redshifts.
Among the new data points we introduce in this work compared to the baseline DES Y3 $3\times2$pt analyses, $w_{s\kappa}(\vartheta)$ is impacted by the relatively large-scale red central galaxy IA, which can be well described by the nonlinear alignment model~\citep[see][]{hs04,brk07,Krause16_IA}.
Compared to $w_{g\kappa}(\vartheta)$, the small-scale $\xi_\pm(\vartheta)$ receives more contamination from satellite alignment, which can not be well described by the NLA model. 

However, we do not expect the tidal alignment and tidal torquing model~\citep[TATT,][]{BMTX19} to outperform NLA in small-scale $\xi_\pm(\vartheta)$ IA modeling.
Recent analyses~\citep{BKC+23,CK24,MAB+24} that compare Effective Field Theory (EFT) of IA and halo alignment in $N$-body simulations find that the TATT model starts breaking down at scales $k\geq 0.05\,h/\mathrm{Mpc}$~\citep[see][for a review]{C25}.
Therefore, we continue to use NLA as our baseline model to mitigate IA in this work and sample both the IA amplitude $A_\mathrm{IA}$ and the redshift evolution power index $\eta_\mathrm{IA}$ with a flat prior between $[-5,\,5]$. 
This choice has been adopted in various DES Y3 re-analyses~\citep[e.g.][]{AAZ+23,ZSM+23,DES_KiDS_WL_23,BAS+24,PH+25,GSJ+25,ZJ25}.
We discuss the impact of using TATT instead of NLA in Sec.~\ref{sec:robustness}.

\subsubsection{Point spread function (PSF) bias} 
\label{sec:sys_psf}
Imperfect PSF subtraction leads to residual PSF ellipticity in the galaxy shape measurement, which translates into biases of $\xi_\pm(\vartheta)$. 
The best-fitting PSF contamination $\delta\xi_\pm^\mathrm{PSF}(\vartheta)$ from DES Y3 shows increasing bias towards smaller scales~\citep{GSA+21,JBA+21,DES_Y3_WL,JOC+25}. 
Given the increased constraining power of our $6\times2$pt analysis, we revisit the bias introduced by PSF contamination in this paper, especially since we are including $\xi_\pm(\vartheta)$ data points below the standard DES Y3 scale cut. 

The estimated shear of galaxies $\bm{\gamma}_\mathrm{est}$ can be contaminated by imperfect PSF modeling $\delta\bm{e}_\mathrm{PSF}^\mathrm{sys}$ and is expressed as 
\begin{equation}
    \bm{\gamma}_\mathrm{est}=\bm{\gamma}+\delta\bm{e}_\mathrm{PSF}^\mathrm{sys}+\delta\bm{e}^\mathrm{noise},
\end{equation}
where $\delta\bm{e}^\mathrm{noise}$ is measurement noise and $\langle\delta\bm{e}^\mathrm{noise}\rangle$ is expected to be zero. 
We note that $\bm{\gamma}_\mathrm{est}$ has accounted for the shear response of \textsc{metacalibration}~\citep{GSA+21,JBA+21}. 
The PSF shear bias term is further decomposed into three contributions~\citep{PAV+08,JSZ+16}
\begin{equation}
    \delta\bm{e}_\mathrm{PSF}^\mathrm{sys} = \alpha\bm{e}_\mathrm{model}+\beta(\bm{e}_*-\bm{e}_\mathrm{model})+\eta\left(\bm{e}_*\frac{T_*-T_\mathrm{model}}{T_*}\right),
\end{equation}
where $\alpha$, $\beta$, $\eta$ are free coefficients per tomography bin, $\bm{e}_\mathrm{model}$ and $T_\mathrm{model}$ are the predicted PSF ellipticity and size from \textsc{Piff}~\citep{JBA+21}, $\bm{e}_*$ and $T_*$ are the true PSF ellipticity and size. For simplicity, we abbreviate those contributions as $\bm{p}\equiv\bm{e}_\mathrm{model}$, $\bm{q}\equiv \bm{e}_*-\bm{e}_\mathrm{model}$, and $\bm{w}\equiv\bm{e}_*(T_*-T_\mathrm{model})/T_*$. 
Assuming that $\bm{p}$, $\bm{q}$, and $\bm{w}$ do not correlate with cosmic shear and galaxy overdensity, the PSF residual only impacts $\xi_\pm(\vartheta)$ in the $6\times2$pt by generating an additive bias
\begin{equation}
    \delta\xi_\pm^{ab,\,\mathrm{PSF}}(\vartheta)=(\alpha_a,\,\beta_a,\,\eta_a)\cdot
    \begin{pmatrix} 
    \rho_0 & \rho_2 & \rho_5 \\
    \rho_2 & \rho_1 & \rho_4 \\
    \rho_5 & \rho_4 & \rho_3
    \end{pmatrix}
    \cdot
    \begin{pmatrix}
        \alpha_b\\
        \beta_b\\
        \eta_b
    \end{pmatrix},
\end{equation}
where the so-called rho-statistics~\citep{Rowe10,JSZ+16} are defined as $\rho_0\equiv\langle\bm{p}\,\bm{p}\rangle$, $\rho_1\equiv\langle\bm{q}\,\bm{q}\rangle$, $\rho_2\equiv\langle\bm{q}\,\bm{p}\rangle$, $\rho_3\equiv\langle\bm{w}\,\bm{w}\rangle$, $\rho_4\equiv\langle\bm{w}\,\bm{q}\rangle$, and $\rho_5\equiv\langle\bm{w}\,\bm{p}\rangle$. We note that each rho statistic has a $+$ and $-$ component as cosmic shear.

Assuming $\bm{p}$, $\bm{q}$, and $\bm{w}$ are the same between galaxies and stars, the rho statistics can be measured from stars. The coefficients $(\alpha,\,\beta,\,\eta)$ of each tomography bin can be measured by fitting the tau statistics
\begin{equation}
\begin{aligned}
\tau_0 & \equiv\langle\bm{\gamma}_\mathrm{est}\,\bm{p}\rangle = \alpha\langle\bm{p}\bm{p}\rangle + \beta\langle\bm{q}\bm{p}\rangle + \eta\langle\bm{w}\bm{p}\rangle,\\
\tau_2 & \equiv\langle\bm{\gamma}_\mathrm{est}\,\bm{q}\rangle = \alpha\langle\bm{p}\bm{q}\rangle + \beta\langle\bm{q}\bm{q}\rangle + \eta\langle\bm{w}\bm{q}\rangle,\\
\tau_5 & \equiv\langle\bm{\gamma}_\mathrm{est}\,\bm{w}\rangle = \alpha\langle\bm{p}\bm{w}\rangle + \beta\langle\bm{q}\bm{w}\rangle + \eta\langle\bm{w}\bm{w}\rangle,\\
\end{aligned}
\end{equation}
using the rho statistics measured from stars. 
An estimation of $\delta\xi_\pm^{ab,\mathrm{PSF}}(\vartheta)$ can be derived given the best-fitting coefficients.

We re-measure the tau- and rho-statistics and run an MCMC chain to derive the best-fitting contamination $\delta\xi_\pm^{\mathrm{PSF,\,bf}}(\vartheta)$. We also calculate $\delta\xi_\pm^{\mathrm{PSF,\,+2\sigma}}(\vartheta)$, which corresponds to the $+2\sigma$-level contamination. 
The $\delta\xi_\pm^{\mathrm{PSF,\,+2\sigma}}(\vartheta)$ is used to contaminate the mock data vectors in Sec.~\ref{sec:scale_cut} for impact study and scale cut decision.

\subsubsection{Nonlinear power spectrum}
\label{sec:sys_nl}
We adopt \textsc{Halofit} as our baseline nonlinear gravity-only matter power spectrum model because of its robustness over a large cosmological parameter space that is relevant to this DES Y3 $\times$ Planck PR4 analysis and because \citep{KFP+21} have shown that \textsc{Halofit} is sufficient for DES Y3 $3\times2$pt analysis. 

However, we re-assess the impact of $P_\mathrm{NL}(k)$ model uncertainties due to the inclusion of small-scale cosmic shear data and the increased cosmological information from CMB lensing and its cross-correlations.

We show the ratio between alternative $P_\mathrm{NL}(k)$ models (\textsc{BACCOEmu}, \textsc{CosmicEmu}, \textsc{EuclidEmu2}, \textsc{HMCode2020}) and the baseline \textsc{Halofit} prediction in Fig.~\ref{fig:Pk_antilles} as the black lines, where thick lines indicate the ratio evaluated within the native $k$ range and thin lines indicate the ratio evaluated from cubic spline-extrapolation of the corresponding $P_\mathrm{NL}(k)$.

We see that different $P_\mathrm{NL}(k)$ models can have 10 percent level difference at $k\geq10\,h/\mathrm{Mpc}$. Although baryonic feedback generates a much stronger impact on $P_\mathrm{NL}(k)$ over this $k$ range, a model mismatch in gravity-only $P_\mathrm{NL}(k)$ can be absorbed by the baryonic feedback PCA model and lead to biased estimation of feedback strength $Q_1$, which translates into biases in $S_8$, $\Omega_\mathrm{m}$.

Therefore, we conduct impact studies of the nonlinear matter power spectrum, and its crosstalk with other systematics, by generating mock data vectors with \textsc{EuclidEmu2}~\citep{EuclidEmu2} and analyzing them using \textsc{Halofit}. 

\subsection{Scale cuts and validation}
\label{sec:scale_cut}
\begin{figure*}
    \centering
    \includegraphics[width=\linewidth]{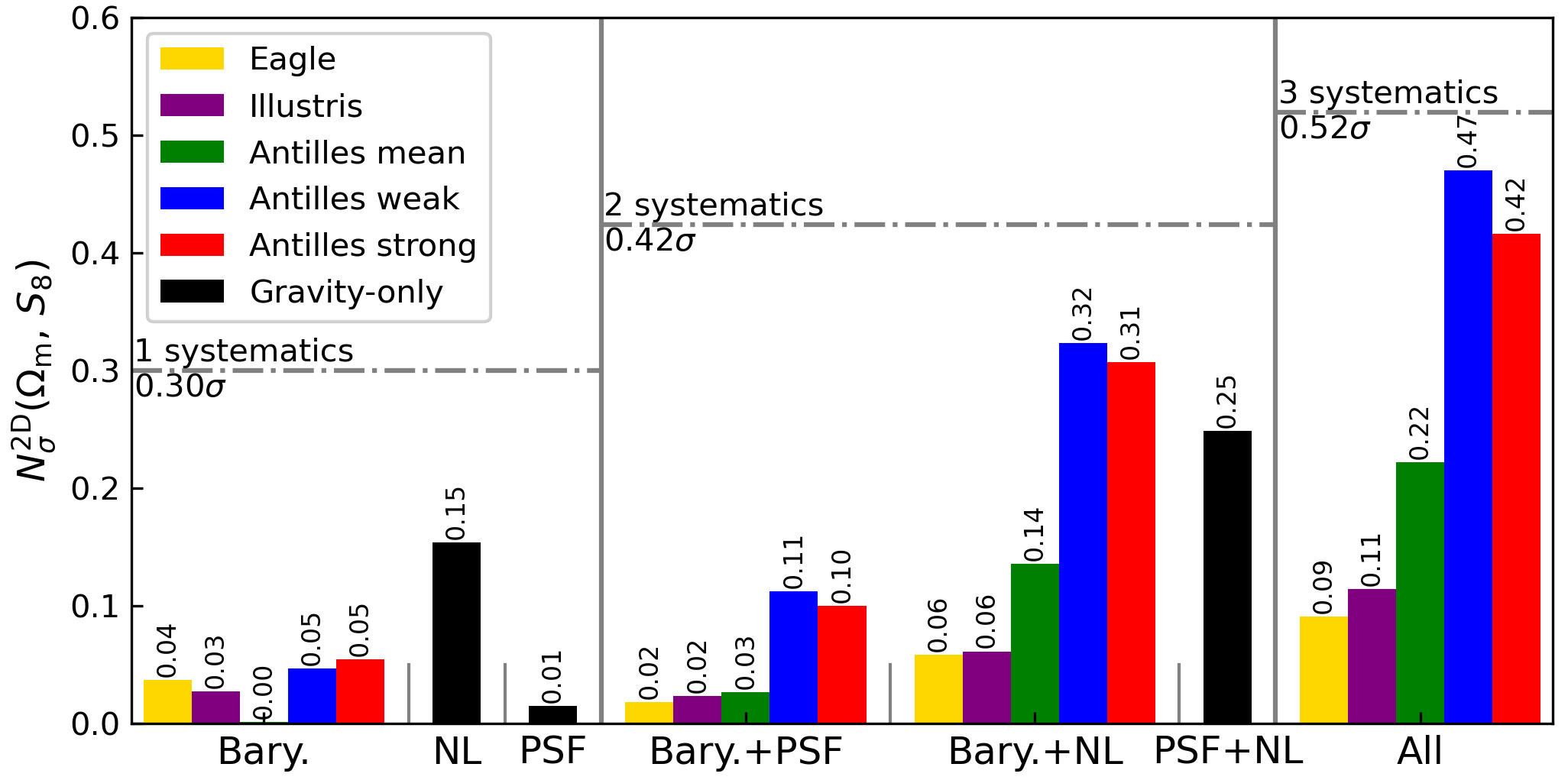}
    \includegraphics[width=\linewidth]{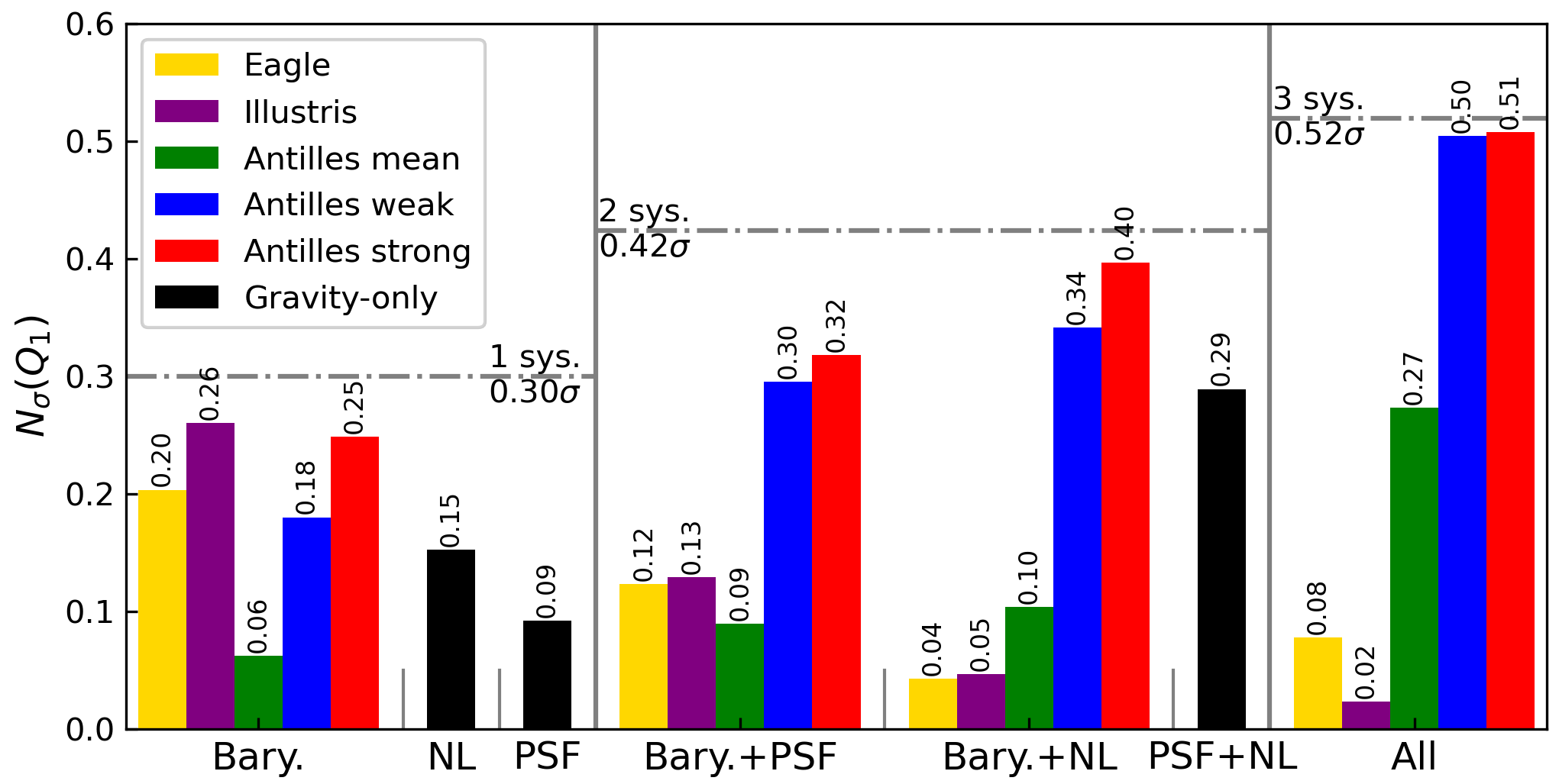}
    \caption{\textit{Top:} 2D bias in $\Omega_\mathrm{m}$-$S_8$ for mock analyses with wide $Q_1$ prior and different systematics in the fiducial data vector. From left to right, we show analyses that include one, two, and three systematics out of baryonic feedback (Bary.), nonlinear matter power spectrum model mismatch (NL), and PSF contamination in $\xi_\pm(\vartheta)$ (PSF). Six baryonic feedback scenarios are considered: gravity-only (black), Eagle (yellow), Illustris (purple), \antilles-strong (red), \antilles-mean (green), and \antilles-weak (blue). We also show the error budget limit for one, two, and three systematics with grey dotted-dashed lines. \textit{Bottom:} Similar to top panel but showing 1D $Q_1$ bias for mock analyses with different systematics in the fiducial data vector.}
    \label{fig:error_budget_250}
\end{figure*}

\begin{figure}
    \centering
    \includegraphics[width=\linewidth]{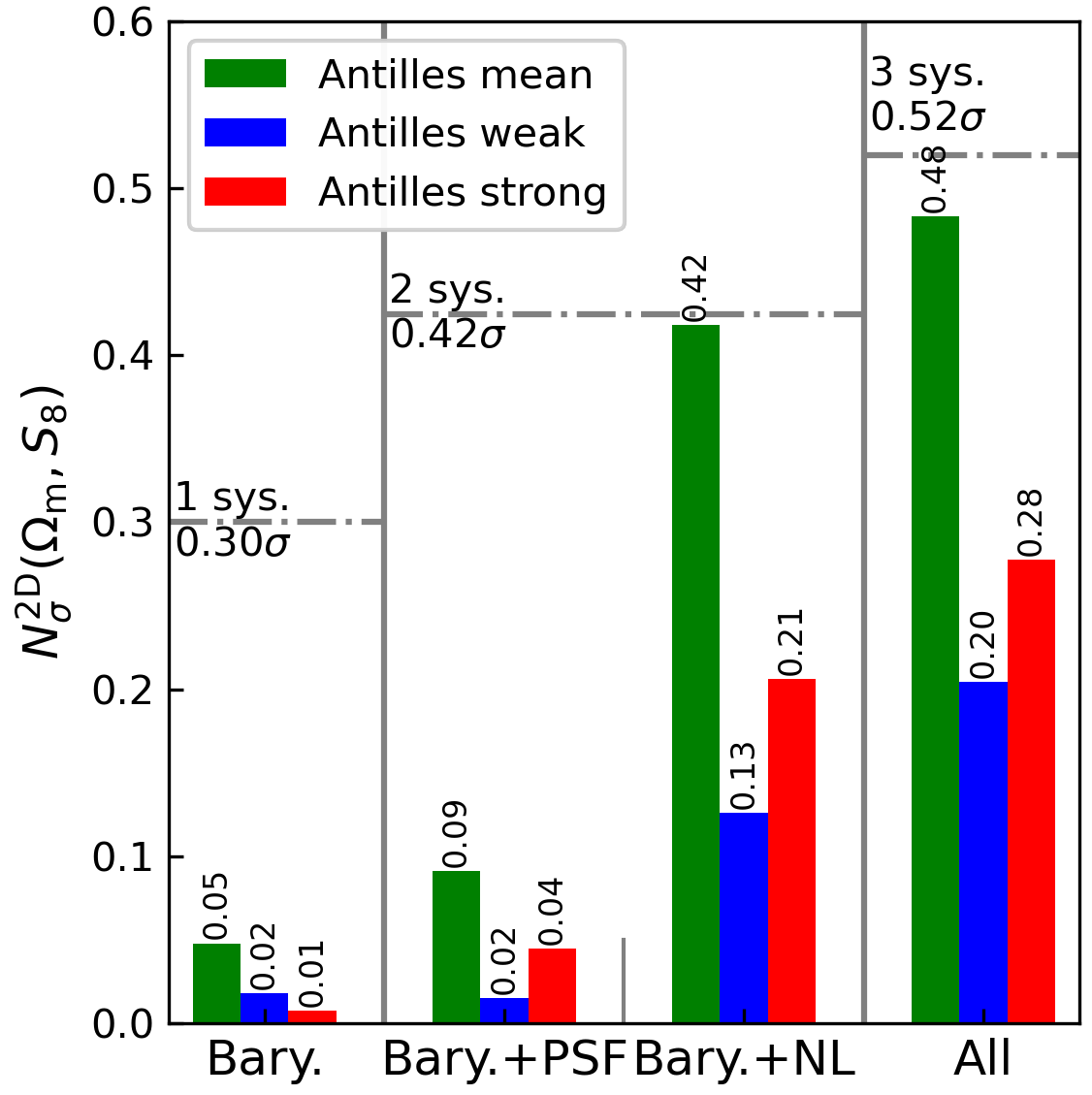}
    \caption{Similar to Fig.~\ref{fig:error_budget_250} but showing 2D $\Omega_\mathrm{m}$-$S_8$ bias for mock analyses including $Q_1$ prior from mean baryon mass fraction observation.}
    \label{fig:error_budget_2D_250fb}
\end{figure}
We include all cosmic shear data points down to $2.^\prime5$ in this work. 
We impose the same scale cuts as~\citep{DES_Y3_3x2pt} in galaxy-galaxy lensing and galaxy clustering (linear galaxy bias). 
\citep{DES_Y3_6x2pt_II_measurement} includes small-scale data points in galaxy-CMB lensing and shear-CMB lensing, which is enabled by the high-fidelity CMB lensing reconstruction from Planck and SPT~\citep{OBC+23}. We choose a conservative scale cut in CMB lensing cross-correlations due to the low angular resolution of Planck PR4 CMB lensing reconstruction. 

The angular scale cut of $w_{g\kappa}(\vartheta)$ is derived from $R_\mathrm{min}=6\,\mathrm{Mpc}/h$, which is the same as $\gamma_t^{ab}(\vartheta)$ in DES Y3~\citep{KFP+21,PCE+22}. 
This is a more conservative choice than~\cite{OBC+23} which adopts $R_\mathrm{min}=3.5\,\mathrm{Mpc}/h$. 
The angular scale cut of $w_{s\kappa}(\vartheta)$ is determined by the Gaussian beam size we applied to the Planck PR4 convergence field. We truncate $w_{s\kappa}(\vartheta)$ at 14 arcmin, which is two times the full-width at half maximum of the Gaussian beam.
We include all $C_{L_\mathrm{b}}^{\kappa\kappa}$ data points in~\citep{CML22}.

We validate our analysis choice by running hundreds of mock analyses using the emulator we trained in Sec.~\ref{sec:emu}. Specifically, we consider the robustness of our pipeline and analysis choices against three modeling uncertainties: baryonic feedback modeling, PSF contamination modeling, and nonlinear matter power spectrum modeling. 

As a metric we require that both the 2D bias in $\Omega_\mathrm{m}$-$S_8$ and the 1D bias in $Q_1$ are lower than $0.3\sigma$, $0.3\sigma\times\sqrt{2}\approx0.42\sigma$, and $0.3\sigma\times\sqrt{3}\approx0.52\sigma$ when 1, 2, and 3 systematics mentioned above are included in the data vector. 
These thresholds were set before the simulated likelihood analyses were conducted. 

We show the $\Omega_\mathrm{m}$-$S_8$ bias (top panel) and the $Q_1$ bias (lower panel) in Fig.~\ref{fig:error_budget_250}. We generate data vectors for six baryonic feedback scenarios: gravity-only, Eagle-like, Illustris-like, \antilles-strong-like, \antilles-mean-like, and \antilles-weak-like contamination.

In all cases, the biases in $\Omega_\mathrm{m}$-$S_8$ and $Q_1$ are below our threshold, although we note that when three systematics are included, the crosstalk between different systematics increases the resulting biases considerably. Not unexpectedly, we observe borderline concerning biases for cases that include both baryonic feedback and nonlinear power spectrum modeling uncertainties.

The biases in $\Omega_\mathrm{m}$-$S_8$ and $Q_1$ for mock \antilles-strong and \antilles-weak data vectors with PSF contamination and $P_\mathrm{NL}(k)$ from \textsc{EuclidEmu2} included are almost surpassing our error budget. This calls for a more precise and robust systematics mitigation method when pushing to smaller scales or when analyzing future, more constraining datasets.

In Fig.~\ref{fig:error_budget_2D_250fb}, we show the 2D $\Omega_\mathrm{m}$-$S_8$ bias similar to Fig.~\ref{fig:error_budget_250}, but consider the case of an additional $Q_1$ prior from $\bar{Y}_\mathrm{b}$ (as we discussed in Sec.~\ref{sec:fb_Q1}). 
In this case, we only consider the three baryonic feedback scenarios for which we can calculate the mean baryon mass fraction in halos of mass between $10^{13}$ and $10^{14}\,M_\odot$: \antilles-strong, \antilles-mean, and \antilles-weak. 
For each of them, we include a $\bar{Y}_\mathrm{b}$-based lognormal $Q_1$ prior centered on their measured $\bar{Y}_\mathrm{b}$ with a standard deviation consistent with equation~\ref{eqn:Q1_fb_prior}.

Comparing Fig.~\ref{fig:error_budget_2D_250fb} to Fig.~\ref{fig:error_budget_250}, we find that biases of \antilles-strong and \antilles-weak are reduced when the $\bar{Y}_\mathrm{b}$-based prior is included. This can be explained by the fact that crosstalk between the nonlinear matter power spectrum, PSF contamination, and baryonic feedback is limited by the tight $Q_1$ prior. 

The only exception is \antilles-mean, whose biases increase due to a biased $Q_1$ prior. We note that the \antilles-mean scenario deviates from the best-fitting $Q_1$-$\bar{Y}_\mathrm{b}$ relation more than the average, hence we consider it to be a stress test for our $\bar{Y}_\mathrm{b}$-based prior method. 

We conclude that the $\bar{Y}_\mathrm{b}$-based $Q_1$ prior methodology is robust to model misspecification even for the most constraining case of a $6\times2$pt analysis.

\section{Cosmology Results}
\label{sec:cosmores}

\subsection{Blinding strategy}
\label{sec:blinding}
Our blinding strategy follows that of~\citetalias{XEM+23}. 
We first validate our \textsc{CoCoA} pipeline by separately reproducing the DES Y3 \textsc{Maglim} $3\times2$pt result~\citep{DES_Y3_3x2pt,PCF+21,DES_Y3_WL} and the Planck PR4 CMB lensing result~\citep{CML22} (see Appendix~\ref{sec:code_compare} for more details).

Real data are only analyzed after successful code comparison, running all simulated likelihood analyses, and after finalizing all analysis choices.

Chains of real data are analyzed after passing the $R-1\leq 0.02$ criterion, where $R$ is the Gelman-Rubin diagnostic. Analyses on real data are also blinded at the posterior level; i.e., our analysis routine only displays posteriors without axis values during blinded analyses. 

Our blinding strategy has three unblinding criteria:
\begin{enumerate}
    \item We conduct consistency tests among different data vector partitions and analysis choices by requiring the 2D marginalized $\Omega_\mathrm{m}$-$S_8$ bias lower than $3\sigma$. 
    The internal consistency analysis routine is designed to test the consistency without disclosing any cosmological parameter values.
    \item For the goodness-of-fit, we compute the reduced $\chi^2/\nu$ of the best-fitting point ($\nu$ is the d.o.f.). We require the probability-to-exceed $\geq 0.01$ given $\chi^2/\nu$.
    \item We visually check that the maxima of the nuisance parameter posteriors are not located near the boundaries of the parameter space in an unexpected way. 
    This test is blind to the actual parameter values. 
\end{enumerate}

\subsection{Weak lensing, galaxy clustering, CMB lensing}
\label{sec:6x2cosmo}

\begin{table*}
    \centering
    \setlength\extrarowheight{2pt}
    \begin{tabular}{llccc}
    \toprule
    Probes & $Q_1$ prior & $S_8$ & $\Omega_\mathrm{m}$ & $\sigma_8$ \\
    \tableline
    \multirow{2}{*}{$6\times2$pt} & $\bar{Y}_\mathrm{b}$ & $0.8073\pm 0.0094$ & $0.292\pm 0.018$ & $0.820\pm 0.025$    \\                   
                          & wide & $0.805\pm 0.012$  & $0.293\pm 0.018$  & $0.816\pm 0.028$    \\
    \tableline
    \multirow{2}{*}{$3\times2$pt} & $\bar{Y}_\mathrm{b}$ & $0.800\pm 0.012$  & $0.302\pm 0.024$  & $0.799^{+0.035}_{-0.041}$   \\
                          & wide & $0.793\pm 0.016$  & $0.310^{+0.027}_{-0.030}$  & $0.782\pm 0.043$   \\
    \tableline
    \multirow{2}{*}{c$3\times 2$pt} & $\bar{Y}_\mathrm{b}$ & $0.822\pm 0.029$  & $0.308^{+0.035}_{-0.044}$  & $0.815^{+0.038}_{-0.042}$   \\
                          & wide & $0.822\pm 0.032$  & $0.307^{+0.038}_{-0.044}$  & $0.816^{+0.038}_{-0.043}$   \\
    \tableline
    \multirow{2}{*}{cosmic shear} & $\bar{Y}_\mathrm{b}$ & $0.787^{+0.020}_{-0.014}$  & $0.255^{+0.037}_{-0.054}$  & $0.863\pm 0.073$\\
                          & wide & $0.786^{+0.022}_{-0.018}$  & $0.256^{+0.038}_{-0.056}$  & $0.860\pm 0.076$ \\
    \tableline
    6$\times$2pt & $\bar{Y}_\mathrm{b}$ & $0.8134\pm 0.0079$ & $0.3057\pm 0.0054$ & $0.8058\pm 0.0069$ \\
     +P1: P-ACT  & wide & $0.8143\pm 0.0086$ & $0.3061\pm 0.0055$ & $0.8063\pm 0.0069$ \\
    \tableline
    6$\times$2pt & $\bar{Y}_\mathrm{b}$ & $0.8086\pm 0.0091$ & $0.3000\pm 0.0064$ & $0.809\pm 0.012$ \\
     +P2: DESI DR2 BAO+BBN & wide & $0.806\pm 0.012$ & $0.3001\pm 0.0069$ & $0.806\pm 0.016$\\
    \tableline
    \multirow{2}{*}{6$\times$2pt + P1 + P2} & $\bar{Y}_\mathrm{b}$ & $0.8079\pm 0.0073$ & $0.2995\pm 0.0034$ & $0.8087\pm 0.0064$ \\
                          & wide & $0.8088\pm 0.0074$ & $0.2996\pm 0.0033$ & $0.8093\pm 0.0067$ \\
    \tableline
    \end{tabular}
    \caption{Summary of the cosmology results for our 6$\times$2pt analysis with different $Q_1$ priors, different data vector partition, and in combination with external probes P-ACT and DESI DR2 BAO+BBN. We show the 1D marginalized constraints on $S_8$, $\Omega_\mathrm{m}$, and $\sigma_8$.}
    \label{tab:cosmology_results_summary}
\end{table*}
\begin{figure*}
    \centering
    \includegraphics[width=0.7\linewidth]{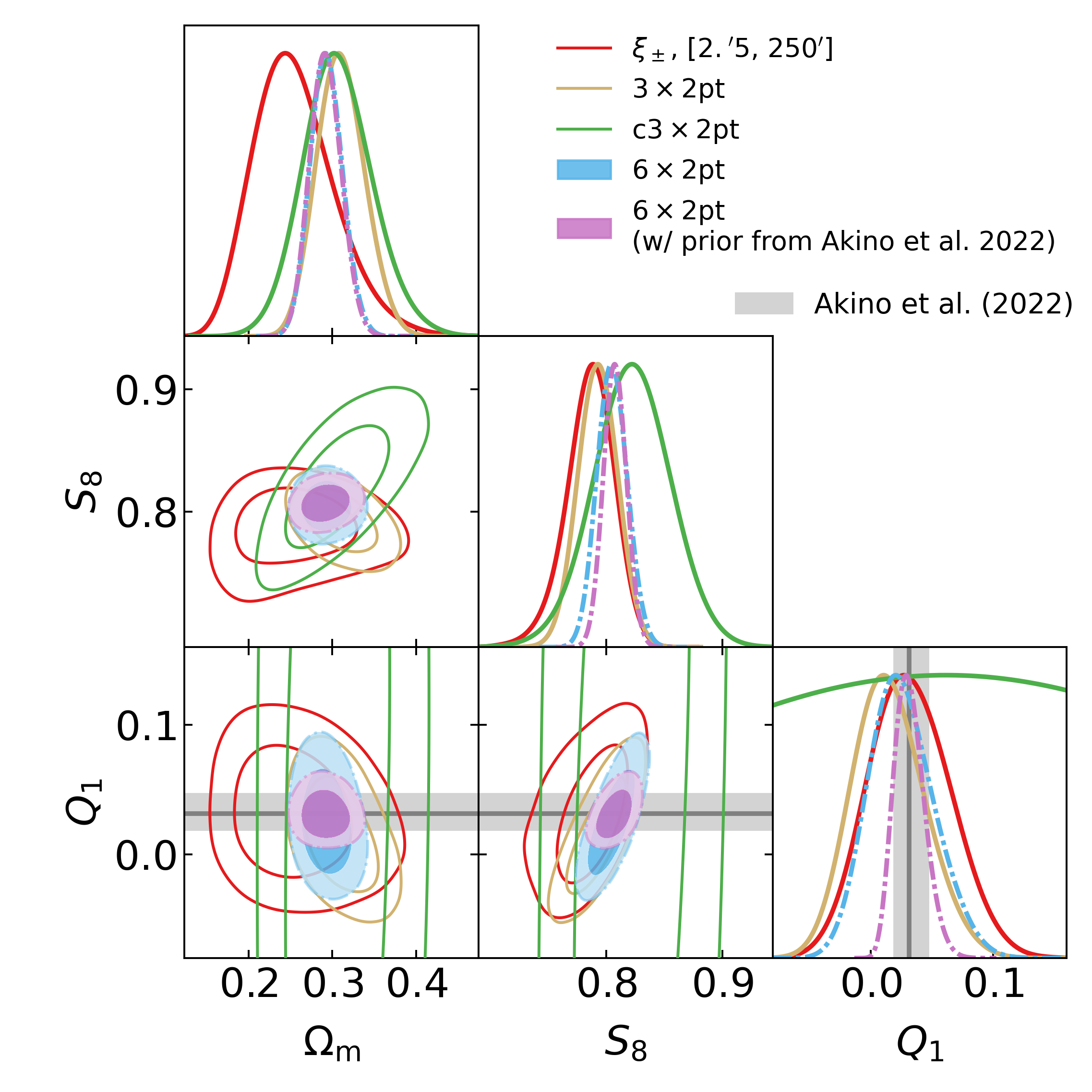}
    \caption{(Color online, real analysis) The results of our baseline 6$\times$2pt analysis with $Q_1$ informed by the $\bar{Y}_\mathrm{b}$ observation in~\cite{AEO+22} (dotted-dashed-filled, purple) compared to results without $Q_1$ prior: cosmic shear (solid-unfilled, red), $3\times2$pt (dashed-filled, golden), c$3\times2$pt (solid-unfilled, green), and $6\times2$pt (dotted-dashed-filled, blue). 
    We show the posterior probability in the $\Omega_\mathrm{m}$, $S_8$, and $Q_1$ and find the subsets of the data vector in good agreement. We show the prediction of $Q_1$ from baryon mass fraction observed in~\citep{AEO+22} as the gray band.}
    \label{fig:res_Y3xP4_6x2pt}
\end{figure*}

\begin{figure*}
    \centering
    \includegraphics[width=0.49\linewidth]{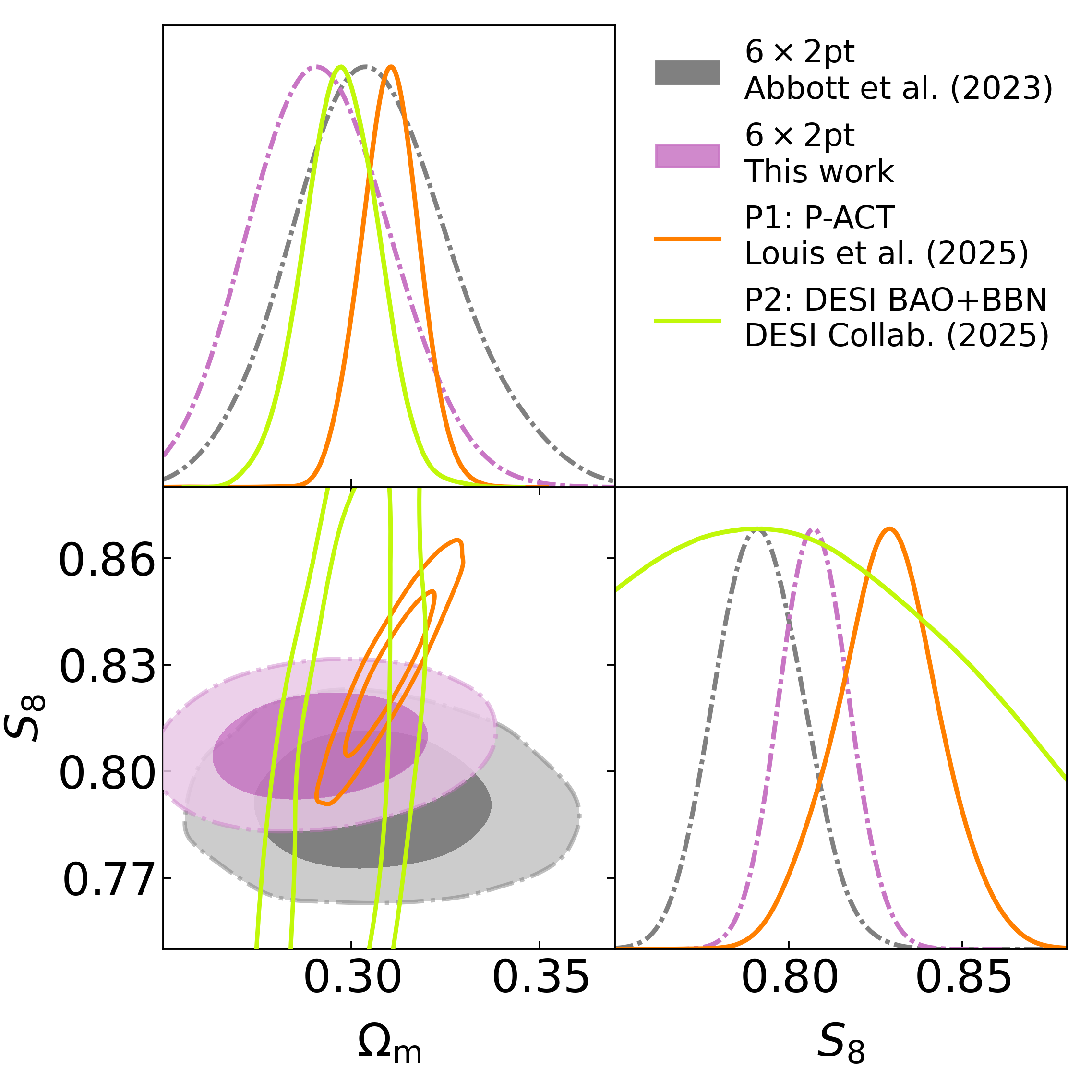}
    \includegraphics[width=0.49\linewidth]{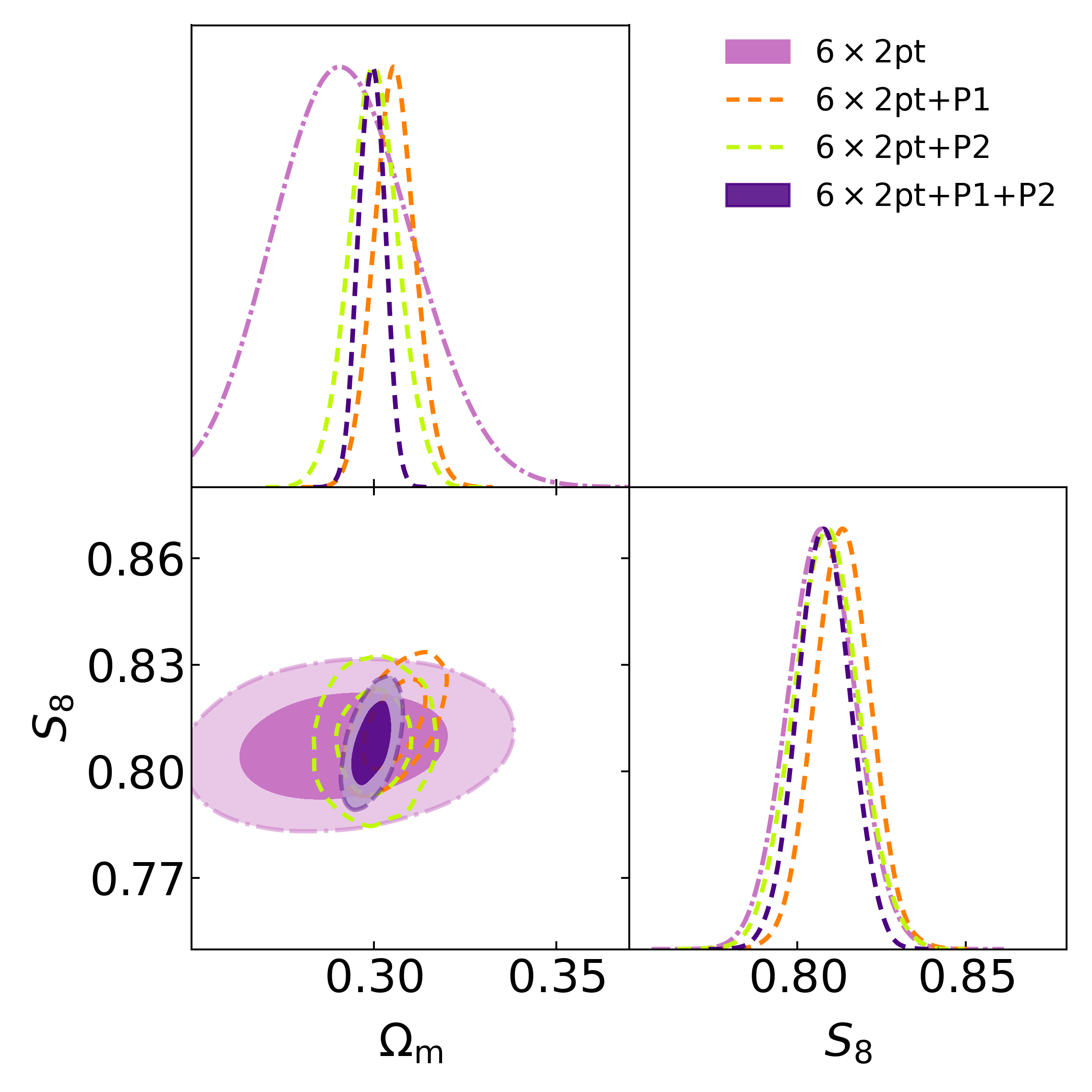}
    \caption{(Color online, real analysis) \textbf{Left panel}: Posterior on $\Omega_\mathrm{m}$-$S_8$ from our $6\times2$pt analysis with $\bar{Y}_\mathrm{b}$-based $Q_1$ prior (purple dotted-dashed-filled contour), the public DES Y3 $\times$ Planck/SPT $6\times2$pt analysis~\citep[gray dotted-dashed-filled contour,][]{DES_Y3_6x2pt_III_23}, joint Planck PR3-ACT DR6 CMB TTEEEE power spectrum (P-ACT, orange solid-unfilled contour), and DESI DR2 BAO+BBN (green solid-unfilled contour). \textbf{Right panel}: Posterior on $\Omega_\mathrm{m}$-$S_8$ for $6\times2$pt analysis with external prior P1 (orange dashed-unfilled contour), P2 (green dashed-unfilled contour), and P1+P2(deep purple dashed-filled contour).}
    \label{fig:6x2pt_DESI_PACT}
\end{figure*}

The MCMC chains on real DES Y3 $\times$ Planck PR4 CMB lensing have passed our unblinding criteria, and we show the corresponding results in Fig.~\ref{fig:res_Y3xP4_6x2pt} and Table~\ref{tab:cosmology_results_summary}.
Different data vector partitions are consistent with each other, and therefore, we are allowed to combine them into $6\times2$pt probes. 

We show the posteriors on $\Omega_\mathrm{m}$, $S_8$, and $Q_1$ from different data vector partitions in our baseline analysis with the wide $Q_1$ prior in Fig.~\ref{fig:res_Y3xP4_6x2pt}. 
When including the $\xi_\pm(\vartheta)$ data down to $2.^\prime5$ and mitigating baryonic feedback with PCA, we derive constraints on $S_8=0.793\pm0.016$ from $3\times2$pt and $S_8=0.805\pm0.012$ from $6\times2$pt. 
We note that the $S_8$ value and error bar are consistent with \cite{DES_Y3_6x2pt_II_measurement}, where the authors discard small-scale $\xi_\pm(\vartheta)$ by adopting the standard DES Y3 scale cut. 
However, we note that \cite{DES_Y3_6x2pt_II_measurement} utilizes a new CMB lensing convergence map reconstructed from Planck and SPT jointly~\citep{OBC+23} and includes small-scale $w_{g\kappa}(\vartheta)$ and $w_{s\kappa}(\vartheta)$. We conclude that including small-scale $\xi_\pm(\vartheta)$ is beneficial. 

Another information from Fig.~\ref{fig:res_Y3xP4_6x2pt} is that the constraint on $Q_1$ from $6\times2$pt alone (the filled blue contour) is consistent with the prediction from \cite{AEO+22} based on the $Q_1$-$\bar{Y}_\mathrm{b}$ relation (the gray shaded region). Therefore, we are allowed to combine the $6\times2$pt likelihood with the $Q_1$ prior from \cite{AEO+22}. 
This leads to a significant improvement on $S_8=0.8073\pm0.0094$, as shown in Fig.~\ref{fig:res_Y3xP4_6x2pt} (the purple filled contour).

\begin{figure*}
    \centering
    \includegraphics[width=0.45\linewidth]{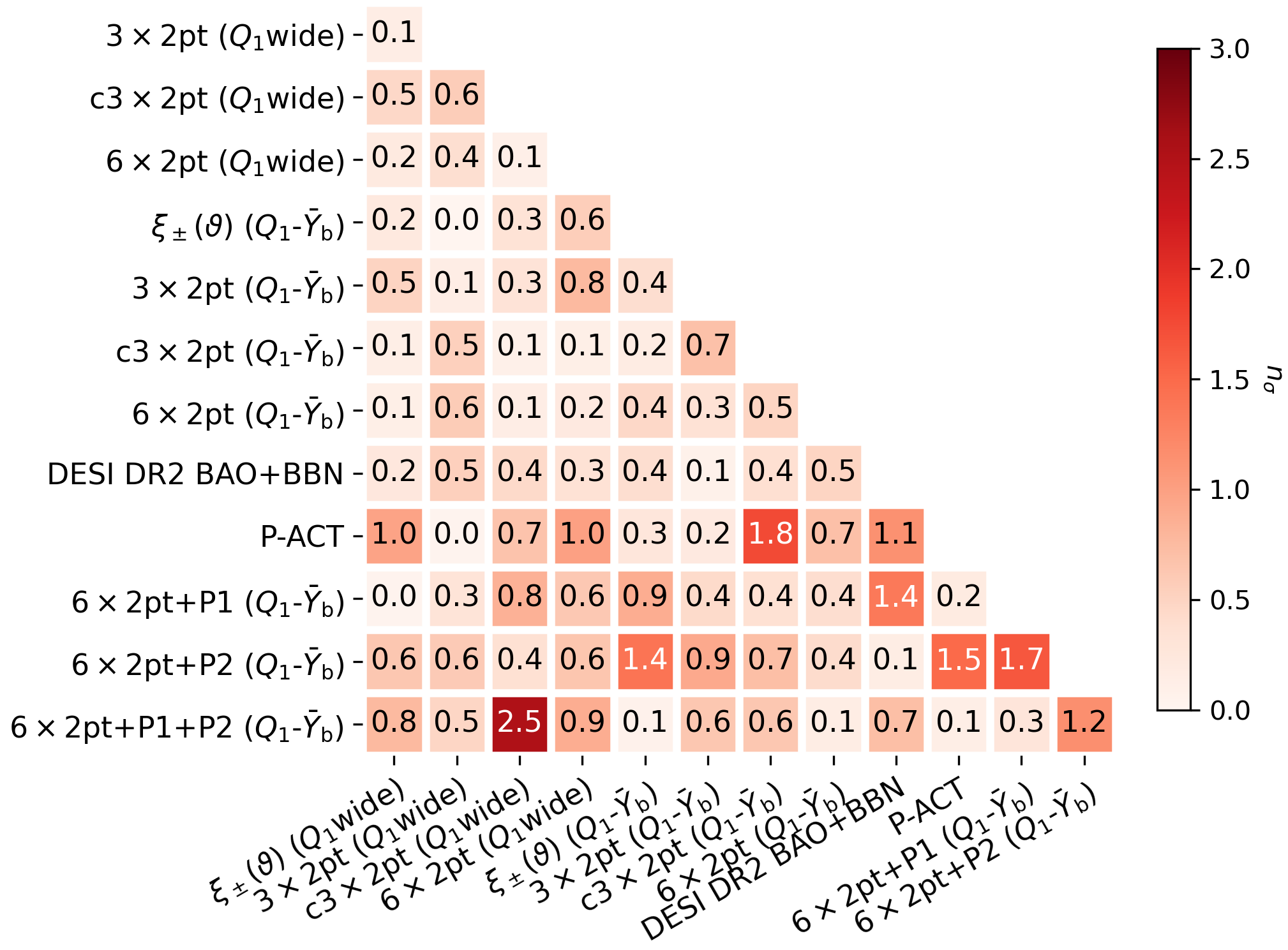}
    \includegraphics[width=0.45\linewidth]{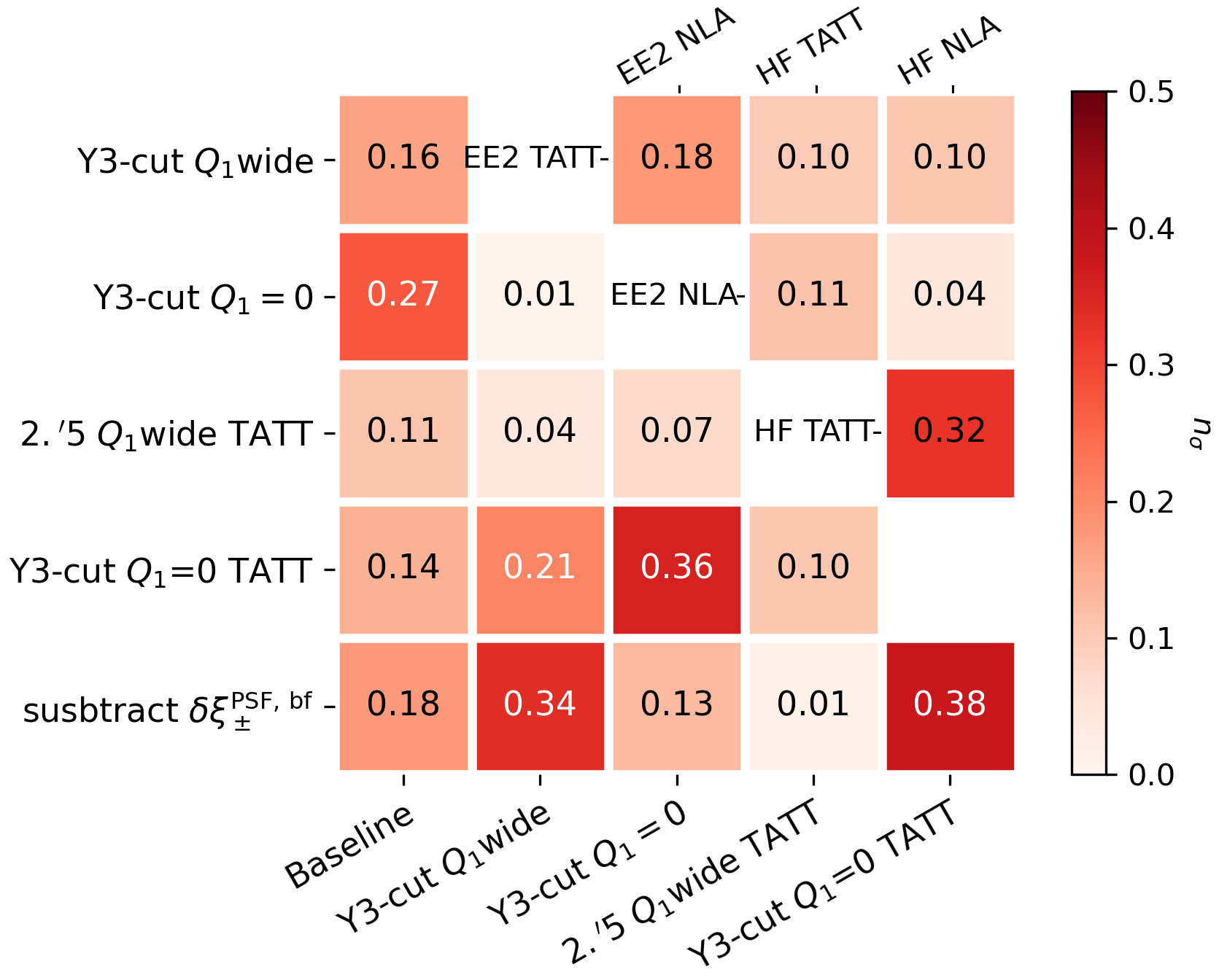}
    \caption{\textbf{Left panel}: tension among cosmological parameters between our $6\times2$pt analysis (and its different partitions) and external probes, including CMB TTTEEE power spectra from Planck PR3 and ACT DR6 (P-ACT, P1) and DESI DR2 all-sample BAO + BBN (P2), and $6\times2$pt in combination with P1 and P2. The heatmap color and the values annotated are the tension metric $n_\sigma$ with the parameter difference method. 
    \textbf{Right panel}: the tension metric $n_\sigma$ among different $6\times2$pt analysis choices. We explain the detailed analysis choices of each row/column in Sec.~\ref{sec:robustness}. The lower-left block of the matrix uses \textsc{Halofit} as the nonlinear matter power spectrum model. The upper-right block further includes a flat prior on cosmological parameters that corresponds to the \textsc{EuclidEmu2} training dataset range for fair comparison between \textsc{Halofit} and \textsc{EuclidEmu2}.}
    \label{fig:tension_metric}
\end{figure*}

We also consider the impact of an alternative $\bar{Y}_\mathrm{b}$ constraint motivated by the eROSITA survey~\citep{PBM+24}, which finds a lower gas mass fraction $f_\mathrm{gas}$ than~\cite{AEO+22}. 
We note that equation~(\ref{eqn:Yb_Q1_fit}) is calibrated from \textsc{ANTILLES} $\bar{Y}_\mathrm{b}$ measurement that includes consistent intracluster light and brightest cluster galaxy mass modeling as~\cite{AEO+22}, which is not available for the GAMA group sample in \cite{PBM+24}. 
Therefore, a rigorous comparison is beyond the scope of this work, and we only estimate the impact of the lower $f_\mathrm{gas}$ found in \cite{PBM+24} assuming a stellar mass fraction from \cite{AEO+22}.
We evaluate $\bar{f}_\mathrm{gas}$ (the mean gas fraction within $[10^{13},\,10^{14}]\,M_\odot$ halo mass) of random realizations of the $f_\mathrm{gas}-M_\mathrm{h}$ relation, which are drawn following equation~(\ref{eqn:fb_scaling_relation}) for \cite{AEO+22}, or their equation (4) for \cite{PBM+24}.
We report $\bar{f}_\mathrm{gas}$ to be $0.0510\pm0.0084$ for \cite{AEO+22} and $0.0427\pm0.0035$ for \cite{PBM+24}, i.e., around $0.9\sigma$ discrepancy (see Fig.~\ref{fig:Q1_fb}).
This $\Delta\bar{f}_\mathrm{gas}$ translates to a change in baryon mass fraction $\Delta\mathrm{log}\left[\bar{f}_\mathrm{b}/(\Omega_\mathrm{b}/\Omega_\mathrm{m})\right]\approx-0.048$, which is within our error budget in equation~(\ref{eqn:fb_Akino_result}). 
Including this $\Delta\bar{f}_\mathrm{gas}$ leads to $S_8=0.8099\pm 0.0096$.

\subsection{Including primary CMB and geometric probes}
\label{sec:6x2cosmo_ext}
Similar to~\citetalias{XEM+23}, we also consider two sets of external cosmological priors:
\begin{enumerate}
    \item P1: Planck PR3+ACT DR6 Primary CMB TTTEEE Power Spectra~\citep[i.e. the P-ACT likelihood in][]{LLA+25}. We are particularly interested in assessing the tension between our $6\times2$pt result and the primary CMB result. We note that the P-ACT likelihood has a derived $S_8=0.830 \pm 0.014$. 
    \item P2: DESI DR2 BAO and Big Bang Nucleosynthesis (BBN)~\citep{DESI_DR2_BAO}. We are interested in this prior as it is geometrical and provides complementary constraints on the expansion history of the Universe. We use the full sample BAO, including the Bright Galaxy Sample, three Luminous Red Galaxies samples, two Emission Line Galaxies samples, the QSO sample, and the Ly$\alpha$ forest sample. The BBN prior is from~\cite{BTV24,S24} and tightly constrain $\Omega_\mathrm{b}h^2=0.02218 \pm 0.00055$.  Combining the DESI DR2 BAO and BBN leads to $\Omega_\mathrm{m}=0.2977 \pm 0.0086$ and $H_0=68.51 \pm 0.58$. Supernovae Ia (SNe Ia) results show $1.7-2.9\sigma$ inconsistency in $\Omega_\mathrm{m}$ with the DESI DR2 BAO+BBN in $\Lambda$CDM, and we do not include them in P2.
\end{enumerate}

We compare the two external priors with our baseline result ($6\times2$pt with $\bar{Y}_\mathrm{b}$ prior) in the left panel of Fig.~\ref{fig:6x2pt_DESI_PACT}, as well as the joint DES Y3 and Planck/SPT $6\times2$pt analysis~\citep{DES_Y3_6x2pt_III_23}. 
\cite{DES_Y3_6x2pt_III_23} is different from our analysis in IA modeling, baryonic feedback mitigation, data vector scale cut, CMB lensing convergence map, covariance modeling, etc. 
Fig.~\ref{fig:6x2pt_DESI_PACT} shows that we obtain lower $\Omega_\mathrm{m}$ and higher $S_8$, but overall, the consistency is good within $0.68\sigma_\mathrm{2D}$ in $\Omega_\mathrm{m}$-$S_8$.

We find good consistency between our baseline result and both P1 and P2 external priors. 
In particular, the 1D $S_8$ marginalized results from our baseline analysis and P-ACT are consistent at $1.35\sigma$.
We evaluate the tension between the baseline result and external priors using a parameter difference-based tension metric~\citep{LRC+21,RD21,RH19,RZH20,DRP+22,SZM+24}, which is also used in \citetalias{XEM+23}. 
We show the tension in Fig.~\ref{fig:tension_metric} left panel and conclude that our baseline $6\times2$pt analysis, P1, and P2 are mutually consistent, and we are allowed to combine them. 

We show the $\Omega_\mathrm{m}$-$S_8$ posterior of $6\times2$pt+$\bar{Y}_\mathrm{b}$ in combination with P1, P2, and P1+P2, on the right panel of Fig.~\ref{fig:6x2pt_DESI_PACT}, and show the 1D marginalized results in Table~\ref{tab:cosmology_results_summary}. 
The most constraining combination $6\times2$pt+$\bar{Y}_\mathrm{b}$+P1+P2 constrains $S_8=0.8079\pm0.0073$ with $0.9$ percent precision, and $\Omega_\mathrm{m}$ constraint is improved significantly (1.1 percent) compared to $6\times2$pt results (6.2 percent). 
The constraining power on $S_8$ saturates with the inclusion of P1 and P2, and adding $\bar{Y}_\mathrm{b}$ prior does not further improve. 

\subsection{Agreement/tensions and robustness tests}
\label{sec:robustness}
\begin{figure}
    \centering
    \includegraphics[width=\linewidth]{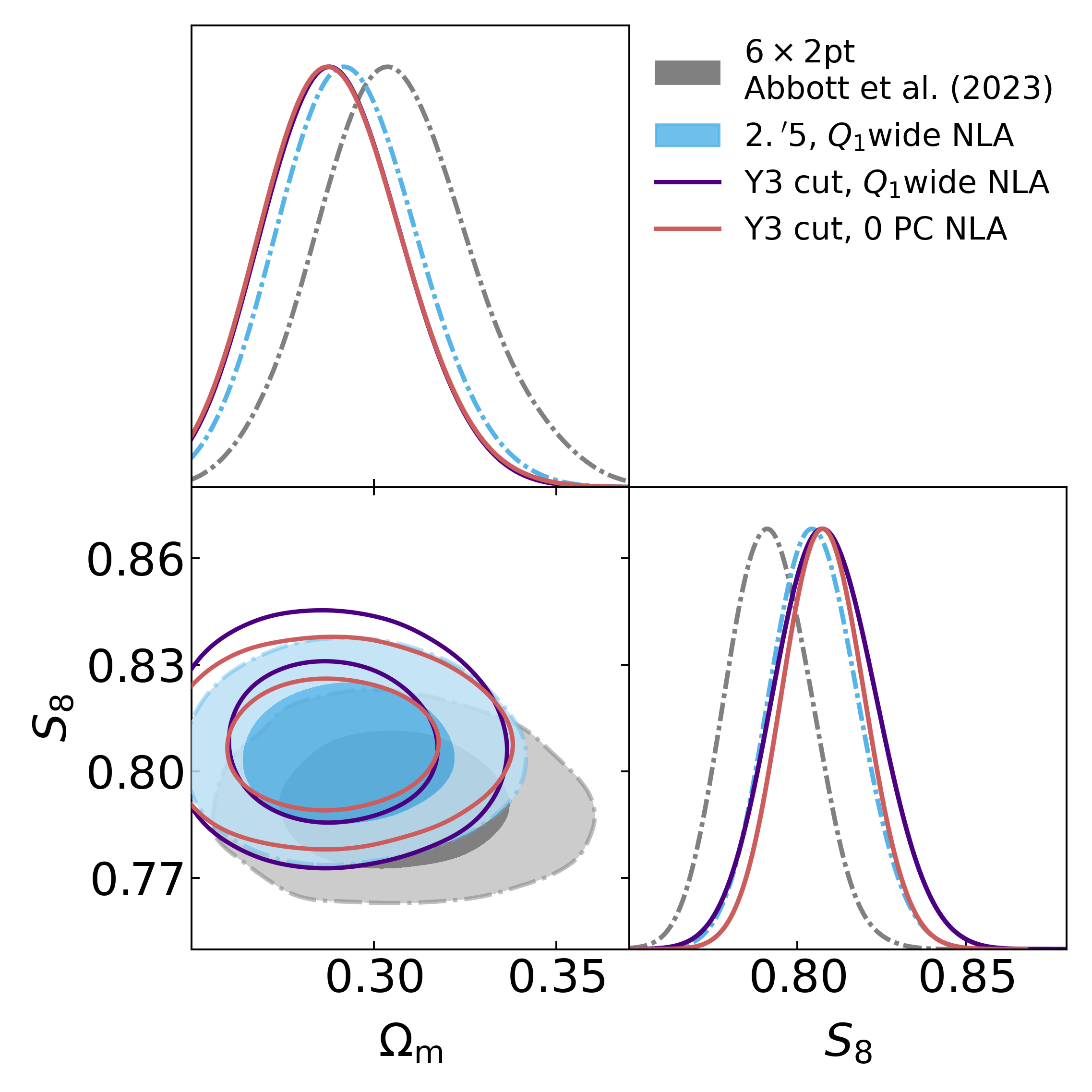}
    \caption{Impact study of different baryonic feedback modeling choices. Posteriors shown in this figure are: the baseline analysis in \cite{DES_Y3_6x2pt_III_23} (the gray filled contour) which adopts TATT and mitigate baryonic impact through scale cut; $6\times2$pt including cosmic shear down to $2.^\prime5$, with a wide $Q_1$ prior (the blue filled contour); $6\times2$pt with DES Y3 $\Lambda$CDM-optimized scale cut and include wide $Q_1$ prior (the purple unfilled contour), $6\times2$pt with DES Y3 $\Lambda$CDM-optimized scale cut and without baryonic feedback PC marginalization (the brown unfilled contour). The three $6\times2$pt posteriors from this work adopt NLA as the IA model.}
    \label{fig:S8_study}
\end{figure}

\begin{figure}
    \centering
    \includegraphics[width=\linewidth]{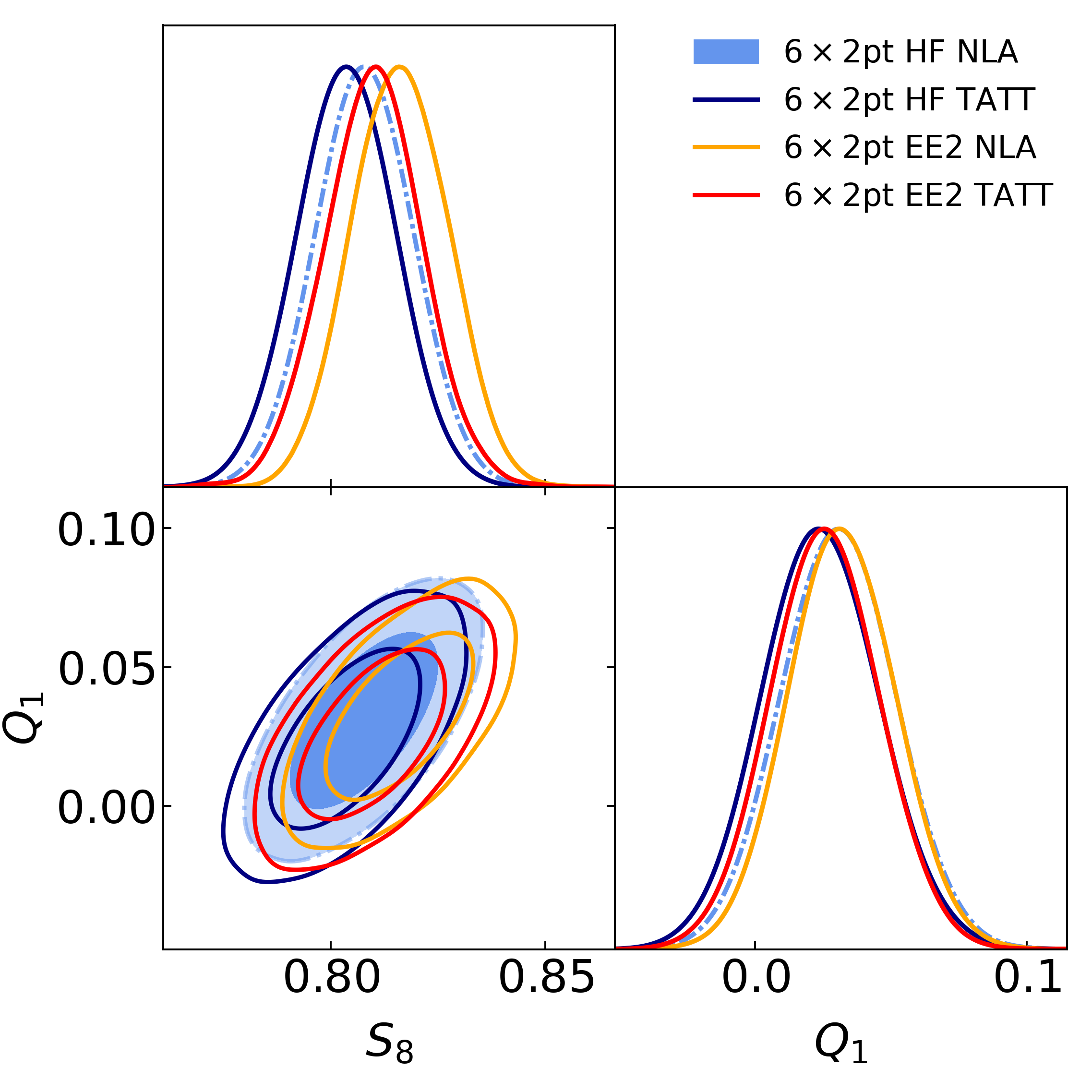}
    \caption{Impact study of different IA models (NLA and TATT) and nonlinear matter power spectra models (\textsc{Halofit}/HF and \textsc{EuclidEmulator2}/EE2). Note that all the chains in this figure are marginalized over $Q_1$ with a wide prior, and include a flat prior on cosmological parameters where the \textsc{EuclidEmulator2} is trained. We annotate the 1D marginalized $S_8$ values in the figure legend. The difference in $\Omega_\mathrm{m}$ is negligible, therefore it is omitted in this figure.}
    \label{fig:S8_study_NL_IA}
\end{figure}

We assess the robustness of our results by running the analysis with alternative analysis choices. We list the considered alternative choices below, sorted by the corresponding systematics:
\begin{itemize}
    \item Baryonic feedback: other than including cosmic shear down to $2.^\prime5$ and sampling the first PC, we also consider the DES Y3 $\Lambda$CDM-optimized $\xi_\pm(\vartheta)$ scale cut, with (\texttt{Y3-cut $Q_1$wide} in the right panel of Fig.~\ref{fig:tension_metric}) or without (\texttt{Y3-cut $Q_1=0$}) marginalizing $Q_1$. 
    
    \item Intrinsic alignment: other than the baseline NLA model, we also consider the TATT model~\citep{BMTX19} as adopted in \cite{DES_Y3_3x2pt}. Particularly, we consider TATT with cosmic shear down to $2.^\prime5$ with wide $Q_1$ prior (\texttt{$2.^\prime5$ $Q_1$wide TATT}) and TATT with the DES Y3 scale cut and no baryonic feedback mitigation (\texttt{Y3-cut $Q_1=0$ TATT}).

    \item PSF contamination: we subtract the best-fitting PSF contamination $\delta\xi_\pm^{\mathrm{PSF,\,bf}}(\vartheta)$ from the measured data vector as the alternative data vector (\texttt{subtract $\delta\xi_\pm^\mathrm{PSF,\,bf}$}).
    
    \item Nonlinear matter power spectrum: other than the baseline \textsc{Halofit}~\citep{SPJ+03,TSN+12,BVH12}, we also consider \textsc{EuclidEmulator2}~\citep{EuclidEmu2} as our nonlinear $P(k)$ model. Note that \textsc{EuclidEmulator2} is valid within a specific parameter space, which is narrower than our $6\times2$pt posterior along dimensions of $A_s$ and $n_s$. Therefore, we impose a flat prior to both \textsc{Halofit} (\texttt{HF NLA} and \texttt{HF TATT}) and \textsc{EmuclidEmulator2} (\texttt{EE2 NLA} and \texttt{EE2 TATT}) chains when comparing between the two models.
\end{itemize}

We show the tension metric $n_\sigma$ in the right panel of Fig.~\ref{fig:tension_metric}. We note that different analysis choices are consistent with each other within $0.4\sigma$, illustrating the robustness of our analysis.

We focus on the impact of baryonic feedback in Fig.~\ref{fig:S8_study}. Comparing among $6\times2$pt analyses down to $2.^\prime5$ with wide $Q_1$ prior (filled blue), with the DES Y3 $\Lambda$CDM scale cut and wide $Q_1$ prior (unfilled purple), and DES Y3 scale cut with $Q_1$ fixed to 0 (unfilled brown), our analyses are consistent among different baryonic feedback mitigation methods. 
This validates the DES Y3 $\Lambda$CDM scale cut as it is immune to baryonic feedback, and extending the analysis down to smaller scales in cosmic shear does not improve the cosmology constraints further, unless an informative prior on the baryonic feedback strength is included from other observations like $f_\mathrm{b}$ or $f_\mathrm{gas}$, X-ray, tSZ, kSZ, fast radio burst dispersion measure, etc~\citep{TMH+22,BAS+24,FAG+24,DTH+25,PKD+22,SAM+25,KNB+25,SKR+25}.

Another message from Fig.~\ref{fig:S8_study} is that, although we do not find strong tension in $S_8$ between our $6\times2$pt analysis and the primary CMB,  baryonic feedback alone does not impact the $S_8$ result significantly,
and is unlikely to be a solution to $S_8$ tension. 
This is similar to part of the conclusions of~\cite{MSS+23,MAS+25}, where the authors find baryonic feedback is not sufficient to resolve the $S_8$ tension. 
Fig.~\ref{fig:S8_study_NL_IA} shows the impact of different IA and nonlinear matter power spectra modeling on $S_8$-$Q_1$ (we omit $\Omega_\mathrm{m}$ due to negligible impact). Compared to NLA, TATT produces lower $S_8$ and requires weaker feedback, and \textsc{Halofit} produces lower $S_8$ than \textsc{EuclidEmulator2}.
The impact of nonlinear $P(k)$ on $Q_1$ is more subtle, with \textsc{EuclidEmulator2} requiring slightly stronger feedback.
The mild differences among different choices (0.26-0.51$\sigma$) indicate that IA and nonlinear $P(k)$ are equally important as baryonic feedback in terms of $S_8$ constraint.
This is consistent with \cite{AAZ+23,MAL24,CDZ+24}, where the authors find that intrinsic alignment, nonlinear matter power spectrum model, and nonlinear galaxy bias can impact the $S_8$ result to a level that is not negligible for $S_8$ tension studies.

Another factor that could bias our result is the dependency of baryonic feedback on cosmology. Intuitively, in a universe where halos are less concentrated, the binding energy is lower, and baryonic feedback can expel gas more efficiently. With a higher mean baryon mass fraction $\Omega_\mathrm{b}/\Omega_\mathrm{m}$, supermassive black holes receive more gas to fuel AGN feedback. This has been found in many studies~\citep{DMS20,SSR+20,SBM+20,AAC+21,DAT+23,EFJ+25}.
\cite{SKD+25} studies the bias on cosmological parameter inference due to cosmology-dependent baryonic feedback strength using the \textsc{Magneticum} multi-cosmology \textsc{Box3/hr} simulation suite, and finds that although the feedback strength correlates strongly with $\Omega_\mathrm{b}/\Omega_\mathrm{m}$ at fixed sub-grid recipe, the PCA-based baryonic feedback mitigation method can recover the fiducial $\Omega_\mathrm{m}$-$S_8$ within $0.3\sigma$ under almost all cosmologies.
Other methods like \textsc{HMCode20} could have $0.5\sigma$ bias under some cosmologies.

\section{Baryonic Feedback Results}
\label{sec:baryonres}

\subsection{Results from this work}
\begin{figure}
    \centering
    \includegraphics[width=\linewidth]{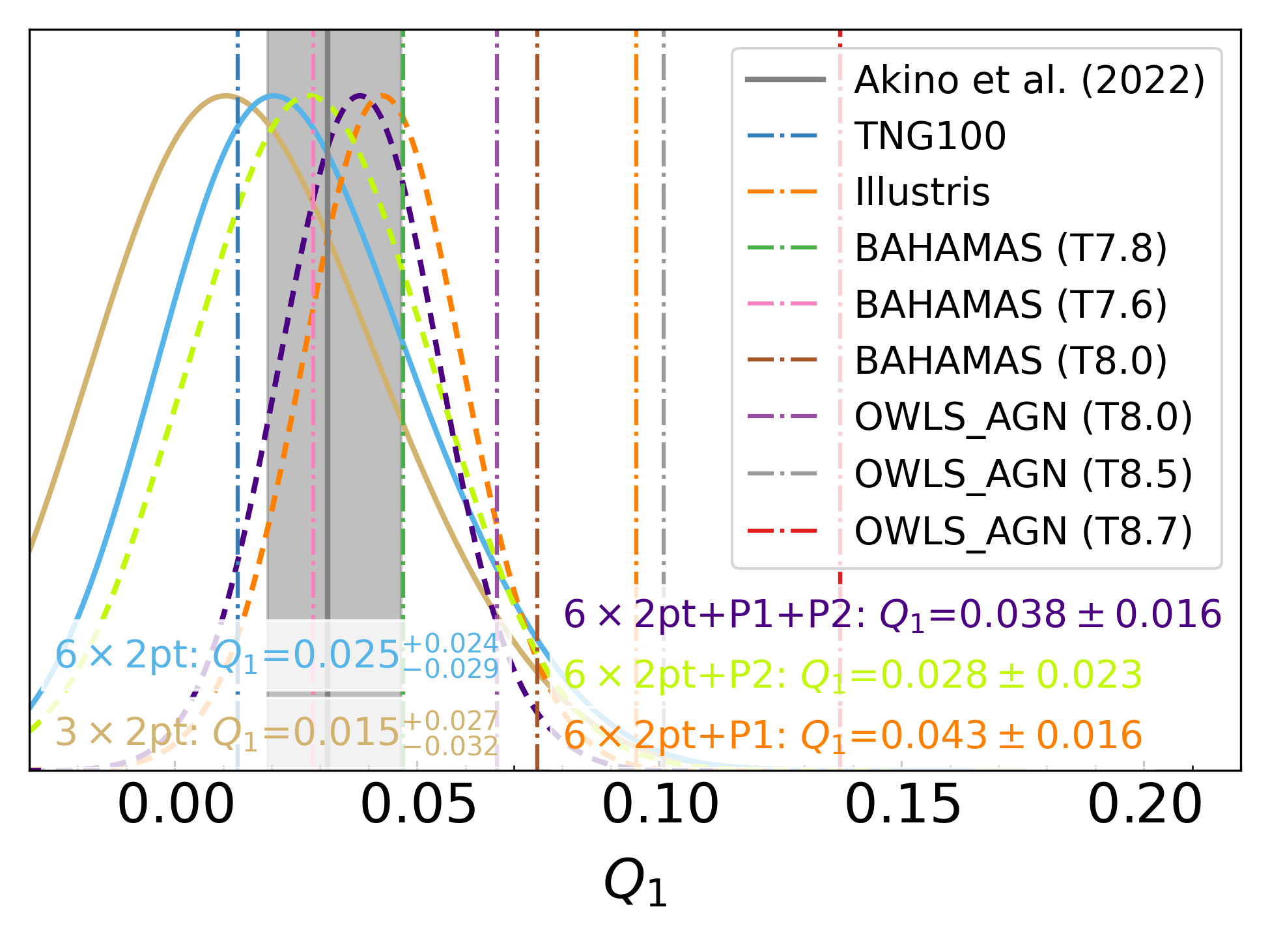}
    \caption{(Color online, real analysis) Posterior on $Q_1$ from our $3\times2$pt (golden), $6\times2$pt (blue), $6\times2$pt+P1 (orange), $6\times2$pt+P2 (light green), and $6\times2$pt+P1+P2 (purple) analysis with wide $Q_1$ prior. We also show the prediction from \cite{AEO+22} in the vertical gray band. $Q_1$ of many hydrosims are shown in vertical dotted-dashed lines.}
    \label{fig:Q1_baseline}
\end{figure}

\begin{figure}
    \centering
    \includegraphics[width=\linewidth]{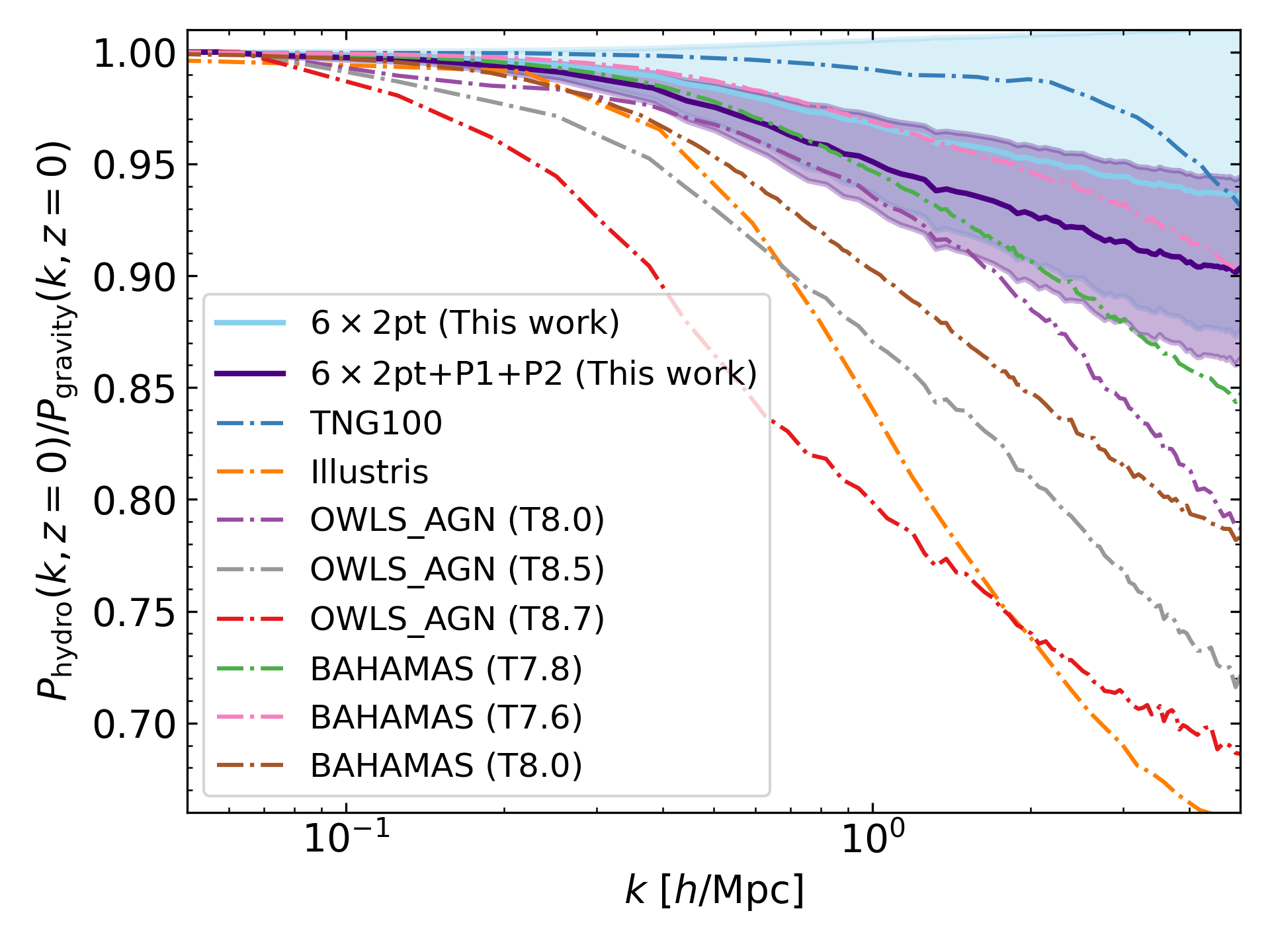}
    \caption{(Color online, real analysis) Posterior on nonlinear matter power spectrum suppression at $z=0$ from our $6\times2$pt (blue) and $6\times2$pt+P1+P2 (purple) analysis with wide $Q_1$ prior. We also show the suppression measured from various hydrosims in dotted-dashed lines. Note that the prediction purely based on $\bar{Y}_\mathrm{b}$ observed in \cite{AEO+22} is very close to the purple shaded region, and is omitted in this figure for clear visualization.}
    \label{fig:Sk_6x2pt}
\end{figure}

We discuss the constraints on baryonic feedback strength $Q_1$ (Fig.~\ref{fig:Q1_baseline}) and the resulting suppression in matter power spectrum (Fig.~\ref{fig:Sk_6x2pt}) in this section. 
All the $3\times2$pt and $6\times2$pt analyses presented in this section (including when combined with CMB or/and BAO priors) include cosmic shear data points down to $2.^\prime5$ and adopt the wide $Q_1$ prior in order to constrain baryonic feedback strength independent from X-ray-based observation~\citep{AEO+22}.
We show the 1D marginalized $Q_1$ constrains in Fig.~\ref{fig:Q1_baseline}, where we find $Q_1=0.015^{+0.027}_{-0.032}$ for the $3\times2$pt analysis and $Q_1=0.025^{+0.024}_{-0.029}$ for the $6\times2$pt analysis.
Including information from external cosmological probes yields slightly stronger feedback: $Q_1=0.043\pm0.016$ ($6\times2$pt + P-ACT), $Q_1=0.028\pm0.023$ ($6\times2$pt + DESI DR2 BAO + BBN), and $Q_1=0.038\pm0.016$ ($6\times2$pt + P-ACT + DESI DR2 BAO + BBN). 
Although we obtain stronger feedback when combined with external probes, the resulting $P(k)$ suppression remains weaker than that of, for example, cOWLS AGN T8.5, and analyses combining with different external probes show excellent consistency in $Q_1$ and $S_8$.

We also show the $Q_1$ of selected hydrodynamical simulations in Fig.~\ref{fig:Q1_baseline}. Compared to our previous DES Y1 $\times$ Planck PR3 $6\times2$pt analysis~\citetalias{XEM+23}, which prefers mild feedback scenarios like BAHAMAS T8.0 and cOWLS-AGN T8.0, our analyses in this work prefer even weaker feedback scenarios like BAHAMAS T7.6, BAHAMAS T7.8, or TNG100. 
Our $Q_1$ results are consistent with the prediction from~\cite{AEO+22}, denoted as the vertical gray band in Fig.~\ref{fig:Q1_baseline}.
Strong feedback scenarios like Illustris ($Q_1=0.095$) and cOWLS AGN T8.7 ($Q_1=0.137$) are excluded at 2.9$\sigma$-3.6$\sigma$ and 4.5$\sigma$-6.2$\sigma$ levels, respectively. 
Our tightest constraint, $6\times2$pt + P1 + P2, when translated to the nonlinear matter power spectrum suppression due to baryonic feedback at $z=0$, indicates $\sim5$ percent suppression around $k=1\,h/$Mpc and $\sim10$ percent suppression around $k=5\,h/$Mpc, as shown in Fig.~\ref{fig:Sk_6x2pt}.
For comparison, $\xi_+^{ab}(2.^\prime5)$ is this work is sensitive to $k\approx0.04$-$0.06\,h/\mathrm{Mpc}$ and $\xi_-^{ab}(\vartheta)$ is sensitive to $k\approx4$-$8\,h/\mathrm{Mpc}$. The small-scale constraint is extrapolated from larger scales via the first PC. 

\begin{figure*}
    \centering
    \includegraphics[width=\linewidth]{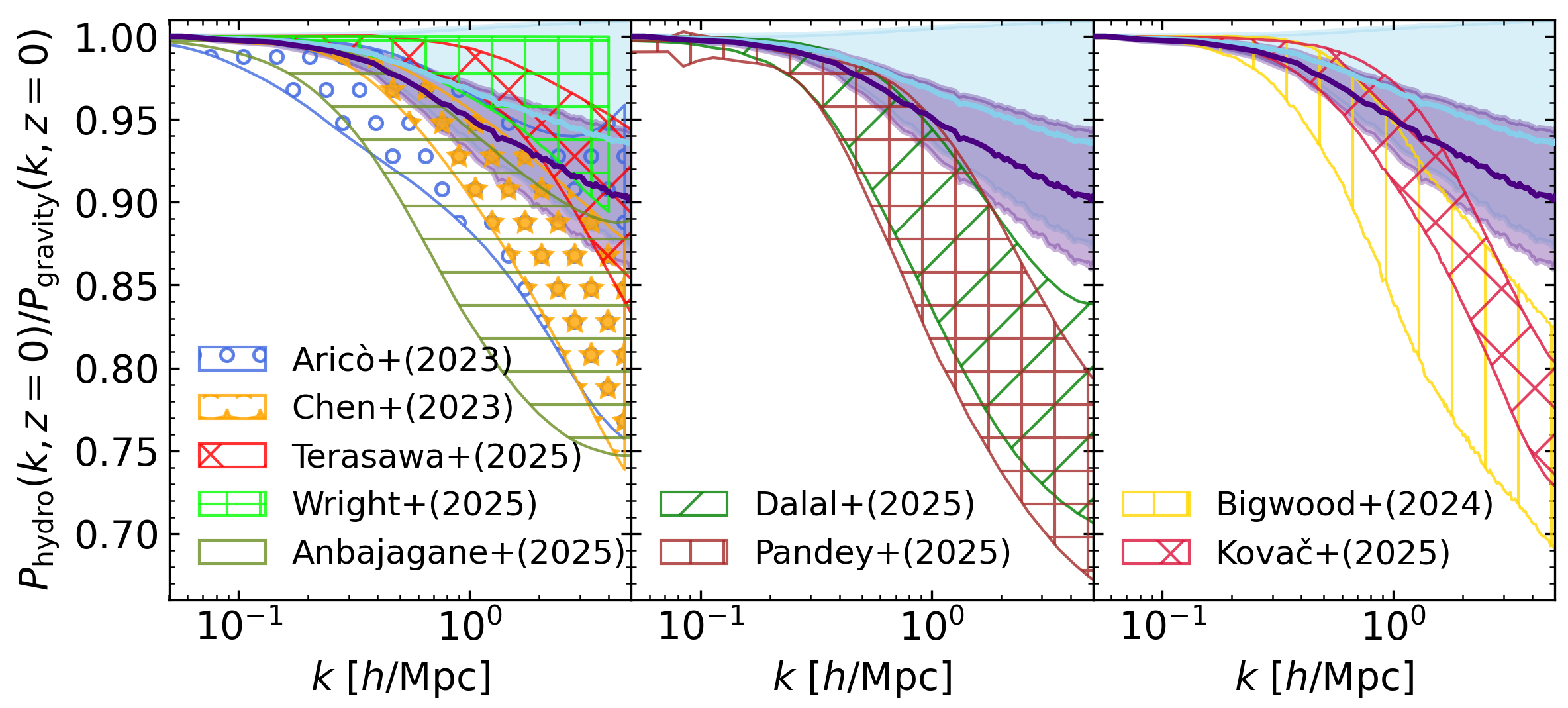}
    \caption{Comparison between the $P(k)$ suppression at $z=0$ from different literature (see Sec.~\ref{sec:Sk_literature}). We show our $6\times2$pt and $6\times2$pt + P1 + P2 results as the light blue and purple solid lines, with the shaded regions showing the $1\sigma$ error budget. Results from the literature are shown as hatched regions, and the boundaries correspond to the 16th and 84th percentiles.}
    \label{fig:Sk_literature}
\end{figure*}

\subsection{Comparison to other literature}
\label{sec:Sk_literature}
We compare our $P(k)$ suppression results to other analyses in literature
\begin{itemize}
    \item \citet{AAZ+23} includes all data points in DES Y3 cosmic shear down to $2.^\prime5$ and analyzes with \textsc{BACCOemu}. The authors constrain the halo mass scale at which halos lose half of their mass $\mathrm{log}M_\mathrm{c}=14.38^{+0.60}_{-0.56}$, which is the only baryonification parameter~\citep{ST15,AAZ+23} that DES Y3 cosmic shear data is sensitive to. 
    
    \item \citet{CAA+23} only uses small-scale cosmic shear data points, which are discarded in the main DES Y3 cosmic shear analysis~\citep{DES_Y3_WL,DES_Y3_WL2}. The authors adopt \textsc{BACCOemu} and constrain $\mathrm{log}M_\mathrm{c}=14.12^{+0.62}_{-0.37}$ with an $\Omega_\mathrm{m}$-$\sigma_8$ prior from DES Y3 $3\times2$pt and other cosmological parameters fixed to Planck 2018 TTTEEE+lowE analysis~\citep{P18A6}.
    
    \item \citet{TLT+25} extends the standard HSC Y3 cosmic shear scale cut~\citep{HSC_Y3_LZS+23} down to $0.^\prime28$ and adopt \textsc{HMCode20}~\citep{MBT+21} to model the baryonic feedback. The baryonic feedback strength is controlled by the temperature of AGN feedback $\Theta_\mathrm{AGN}\equiv\mathrm{log}_{10}(T_\mathrm{AGN}/K)$, with a flat prior of flat[7.3, 8.3].
    
    \item \citet{WSA+25} presents the KiDS Legacy Survey analysis with \textsc{HMCode20} as the baryonic feedback modeling code, and sample $\Theta_\mathrm{AGN}$ with flat prior flat[7.3, 8.3].
    
    \item \citet{ACD+25} uses the cosmic shear 2PCFs measured from the Dark Energy Camera All Data Everywhere (DECADE) project and DES Y3 to constrain both cosmology and baryonic feedback. The authors model baryonic feedback with \textsc{BaryonForge}~\citep{APC24}, which adopts a halo-model formalism similar to that of \cite{MBT+21,PH+25}. \textsc{BaryonForge} includes a variety of dark matter and baryon density profiles, including profiles from \textsc{BCemu}~\citep{STS+19,GS21}, \textsc{BACCO}~\citep{AAC+21}, and \textsc{HMx}~\citep{MBT+21}. We consider the constraints from \textsc{BaryonForge} with \textsc{BCemu} profile presented in \cite{ACD+25}.
    
    \item \citet{DTH+25} applies the Dark Matter + Baryon (DMB) model developed in \cite{TPK+24} to the tSZ $Y$-$M$ relation measured from a tSZ-selected galaxy clusters. The tSZ $y$ signal is measured from ACT DR5, and the hydrostatic mass is measured through weak lensing from DES Y3. The authors then fix cosmology to Planck~\citep{P18A6} and sample DMB parameters, from which the matter power spectrum suppression can be calculated.

    \item \citet{PH+25} conducts joint analysis of DES Y3 cosmic shear 2PCF and cross-correlation between cosmic shear and tSZ $y$ map from ACT DR6 and Planck~\citep{CMD+24}. The authors apply the framework developed in \cite{PST+25} and constrain cosmological and astrophysical parameters that control baryonic feedback.

    \item \citet{BAS+24} jointly analyze the full-scale DES Y3 cosmic shear 2PCF and the stacked kSZ profile measured from BOSS CMASS $\times$ ACT DR5~\citep{SFA+21}. The authors utilize the \textsc{BCemu} model to constrain both cosmology and baryonic feedback.

    \item \citet{KNB+25} jointly analyze the stacked kSZ profile from BOSS CMASS $\times$ ACT~\citep{SFA+21} and halo X-ray gas fractions from eROSITA and pre-eROSITA datasets~\citep{PBM+24,GS21} using the new component-wise baryonification~\citep{SKB+25}. The authors fix cosmology to the fiducial one used in the \textsc{FLAMINGO} simulations and sample the baryonification parameters.
\end{itemize}

Compilation of the power spectrum suppression from the literature above and the result of our analysis are shown in Fig.~\ref{fig:Sk_literature}. We sort the literature results so that $P(k)$ suppression from weak lensing-based analyses (left panel), tSZ-based analyses (middle panel), and kSZ/X-ray-based analyses (right panel) are shown separately. 
We note that our constraints are generally well aligned with the weak lensing-based results, although with mild differences from \cite{ACD+25,CAA+23}. Analyses based on tSZ/kSZ/X-ray generally prefer stronger feedback at $k\geq1 h/\mathrm{Mpc}$. 
We note that strong baryonic feedback is also found in \cite{TMH+22,HFR+25,HFF+25,RSH+25,SAM+25}.

\cite{LPP+25} proposes a potential avenue to resolve this difference by examining the impact of baryonic feedback from a halo assembly perspective: while weak lensing, tSZ, kSZ, and X-ray observations are sensitive to different scales and redshift, the efficiency of baryonic feedback is also a function of halo mass, radial scale, and redshift.  
Massive halos (which weak lensing, tSZ, and X-ray are sensitive to) are impacted by baryonic feedback early, but are affected less, and re-accrete the ejected matter at lower redshifts due to their deep gravity potential. 
Cosmological probes sensitive to massive halos at different redshifts may capture different feedback strengths during halo assembly. 
Feedback in less massive halos (which kSZ is sensitive to) starts at lower redshifts, but can efficiently move gas outside of halos and maintain strong feedback suppression.
However, this simple explanation has difficulty in explaining the strong feedback found in \cite{PH+25}, which is sensitive to massive halos, and the consistency between the DES Y3 shear-only result and the shear+kSZ result in \cite{BAS+24}. Also, the halo mass and redshift range of kSZ measurements overlap with the HSC-XXL X-ray sample considerably, but the kSZ prefers much stronger feedback than the HSC-XXL sample\cite{MAS+25,SAM+25}. 
Therefore, it is possible that there are unknown systematics in some of these measurements that produce the differences in feedback strength measured by different probes. 
More powerful data (e.g. DES Y6, LSST Y1, and Simons Observatory) are required to draw a solid conclusion on $P(k)$ suppression measured by different probes. 

\cite{PPA+25} examines popular analytical models for baryonic feedback~\citep{STS+19,MBT+21,MTH+20,AA24} and finds that, although those models can fit the matter-matter, matter-pressure, and electron density-electron density power spectra with high accuracy, their prediction on halo properties (e.g., bound gas fraction, electron pressure, temperature, and density profiles) varies considerably. This indicates that the current baryonic feedback models need further refinement to model weak lensing and astrophysical probes consistently.
Another recent analysis~\citep{ACD+25} investigates the best-fitting feedback-induced $P(k)$ suppression with the DECADE weak lensing data. The authors consider \textsc{BCEmu}, \textsc{BACCOEmu}, \textsc{HMx}, and \textsc{BaryonForge}, and find that astrophysical and cosmological (e.g. $\Omega_\mathrm{b}/\Omega_\mathrm{m}$) priors explicitly or implicitly assumed in the analysis can impact the $P(k)$ suppression by a couple of percents, indicating a nontrivial projection effect in $P(k)$ suppression analysis.

\section{Conclusions}
\label{sec:conclusion}

In this paper, we constrain cosmology and baryonic physics scenarios with a combination of weak lensing, galaxy clustering, CMB lensing, and cross-correlations using DES Year 3 and Planck PR4 datasets. Compared to previous work (\citetalias{XEM+23}) we have significantly upgraded our catalog-to-cosmology pipeline, specifically in three different areas: 

\begin{enumerate}
\item We improved our baryonic physics modeling by employing 400 \antilles\ hydrodynamical simulations.
\item We derive a scaling relation between the baryonic feedback amplitude parameter $Q_1$ and the mean baryon mass fraction in halos of $M_{500}\in[10^{13},10^{14}]\,M_\odot$. This allows us to incorporate external information \cite{AEO+22} on baryonic physics into our analysis.
\item We train a neural network emulator to run hundreds of simulated analyses exploring the science return as a function of analysis choices. 
\end{enumerate}

The latter is of particular importance for future science analyses in this area. Our emulator generates $6\times2$pt model vectors given input cosmological and nuisance parameters accurately over a large parameter space. This accelerates our $6\times2$pt analysis runtime CPU hours by more than a factor of 10$^3$, allowing us to quickly explore different analysis choices and conduct impact studies for various systematics and their permutations. This ability not only reduces the carbon footprint of our analysis but also ensures the robustness of the pipeline and analysis choices before applying it to real data. 

We studied the impact of three systematics specifically: baryonic feedback, PSF contamination, and nonlinear matter power spectrum model mismatch. 
We conclude that, at the precision of the DES Y3 $\times$ Planck PR4 CMB lensing $6\times2$pt analysis, all systematics are controlled within the previously defined error budget. However, our analysis suggests that a more accurate and robust systematics mitigation method is essential for future DES Y6 $6\times2$pt analyses, in particular when pushing $\xi_\pm(\vartheta)$ below $2.^\prime5$. 

After an extensive suite of synthetic analyses, we obtained competitive results on $S_8$ from our $6\times2$pt analysis, $S_8=0.8073\pm0.0094$ when including the information from \cite{AEO+22}. Our result is consistent with \cite{DES_Y3_6x2pt_II_measurement}.

Our $6\times2$pt is consistent with both the primary CMB TTTEEE spectra from Planck PR3 and ACT DR6 (P-ACT, external prior P1) and the DESI DR2 BAO and BBN (external prior P2). Combined our $6\times2$pt analysis with P1 and P2 tightly constrain $S_8=0.8079\pm0.0073$ and $\Omega_\mathrm{m}=0.2995\pm0.0034$.

We quantify small-scale matter power suppression due to baryonic feedback via the amplitude of the first principal component $Q_1$ in our analysis. We find a preference for weak feedback scenarios ($Q_1=0.025^{+0.24}_{-0.029}$ for $6\times2$pt and $Q_1=0.038\pm0.016$ for $6\times2$pt+P1+P2), which is in strong disagreement with the feedback scenarios of OWLS AGN T8.7 or Illustris simulations. When combined with P1 and P2, we find $\sim 5$ percent suppression around $k\approx1\,h/$Mpc and $\sim 10$ percent suppression at $k\approx 5\,h/$Mpc.

This result is consistent with other weak lensing-based constraints on the small-scale power suppression, but generally weaker than results from tSZ/kSZ/X-ray-based analyses.

Several possible extensions of this work are worth mentioning. On the data side, the near-future DES Year 6 and slightly later LSST Year 1 data are fascinating prospects. On the CMB side, the ACT DR6 dataset exists already, and near-future data from the Simons Observatory poses an exciting opportunity. On the methodology side, the rapidly evolving \textsc{CoCoA} pipeline has fully adopted the machine-learning-accelerated inference concept, which will be even more important as the model space expands beyond $\Lambda$CDM and into more complex systematics parameterizations.

\begin{acknowledgments} 
JX would like to thank Chun-Hao To, Anna Porredon, and Shivam Pandey for sharing DES Y3 multi-probe chains with alternative analysis choices for \textsc{CoCoA} code validation. 
JX thanks Daayha Anbajagane, Giovanni Aricò, Leah Bigwood, Nihar Dalal, Michael Kovač, Shivam Pandey, Ryo Terasawa, and Angus Wright for sharing their matter power spectrum suppression due to baryonic feedback. 
This work is supported by the Department of Energy Cosmic Frontier program, grant DE-SC0025993 and by the “Maximizing Cosmological Science with the Roman High Latitude Imaging Survey” Roman Project Infrastructure Team (NASA grant 22-ROMAN11-0011).
This work was supported by the Science and Technology Facilities Council (grant number ST/Y002733/1). This project has received funding from the European Research Council (ERC) under the European Union’s Horizon 2020 research and innovation programme (grant agreement No 769130).
Simulations in this paper use High Performance Computing (HPC) resources supported by the University of Arizona TRIF, UITS, and RDI and maintained by the UA Research Technologies department.
\end{acknowledgments}

\onecolumngrid
\appendix

\section{Code Comparison}
\label{sec:code_compare}
\begin{figure}
    \centering
    \includegraphics[width=0.4\textwidth]{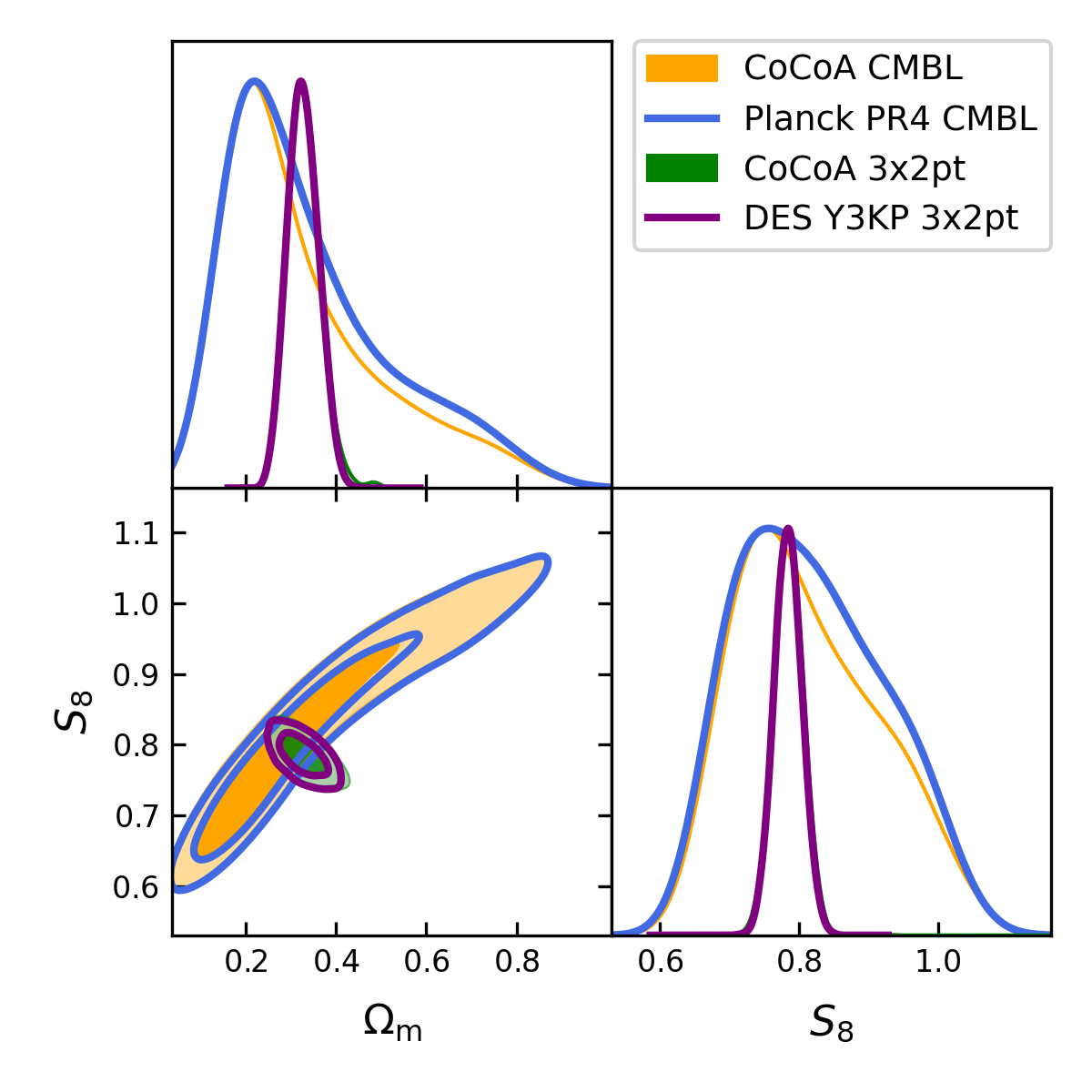}
    \caption{Here we re-analyze the DES Y3 $3\times2$pt of the \textsc{Maglim} sample using \textsc{CoCoA} (green filled contour) and compare the $\Omega_\mathrm{m}$-$S_8$ posterior to the DES Y3 standard results (purple unfilled contour). Both chains assume NLA and do not include shear ratio. We also re-analyze the Planck PR4 CMB lensing band-power using \textsc{CoCoA} (orange filled contour) and compare to the public Planck PR4 result (blue unfilled contour).}
    \label{fig:cocoa_code_comparison}
\end{figure}

In Fig.~\ref{fig:cocoa_code_comparison}, we reproduce the DES Y3 \textsc{Maglim} sample $3\times2$pt result excluding the shear ratio, which is not included in \textsc{CoCoA} pipeline.
We also validate the pipeline by reproducing the result of Planck PR4 CMB lensing~\citep{CML22}.

\twocolumngrid
\bibliography{main}

\end{document}